\documentclass[aps,prl,graphicx,9pt,twocolumn,superscriptaddress]{revtex4-2}
\usepackage{amsmath,amscd,graphicx,graphicx,cases}
\usepackage{multirow,float,subfigure,appendix}
\usepackage{diagbox,makecell,blindtext,color,amssymb}
\usepackage[colorlinks,linkcolor=blue,urlcolor=blue,
anchorcolor=blue,citecolor=blue]{hyperref}
\usepackage{textcomp,booktabs,colortbl,easyReview}
\newcolumntype{C}{c<{\kern\tabcolsep}@{}}
\usepackage{times}
\usepackage{lineno}
\usepackage{xcolor,ulem,bm}
\usepackage{orcidlink}

\begin{document}
\draft
\title{Observation of the Transition from Reversible to Irreversible Decoherence of Mesoscopic Quantum Superpositions}

\author{Ri-Hua Zheng\,\orcidlink{0000-0002-1944-1573}}
\author{Jia-Hao L\"{u}}
\author{Fan Wu}
\author{Yan Xia\,\orcidlink{0000-0002-4539-298X}}\email{xia-208@163.com}
\affiliation{Fujian Key Laboratory of Quantum Information and Quantum Optics, College of Physics and Information Engineering, Fuzhou University, Fuzhou, Fujian, 350108, China}

\author{Li-Hua Lin}
\author{Zhen-Biao Yang\,\orcidlink{0000-0002-3964-4714}}\email{zbyang@fzu.edu.cn}
\author{Shi-Biao Zheng}\email{t96034@fzu.edu.cn}
\affiliation{Fujian Key Laboratory of Quantum Information and Quantum Optics, College of Physics and Information Engineering, Fuzhou University, Fuzhou, Fujian, 350108, China}
\affiliation{Hefei National Laboratory, Hefei 230088, China}

\begin{abstract}
The decoherence of superpositions of classically distinguishable states (cat states) is crucial for understanding quantum-to-classical transitions and quantum measurements. So far, irreversible decoherence processes of mesoscopic cat states have been demonstrated in several experiments. We here report on an experimental demonstration of the dynamics of such states in a reservoir with controllable degrees of freedom, which allows observation of the transition from reversible to irreversible mesoscopic decoherence. The experiment is performed with a circuit quantum electrodynamics device, where a bus microwave resonator storing a photonic cat state is connected to many nonlinear electronic oscillators. Each of the oscillators that are effectively coupled to the bus resonator serves as one degree of freedom of the reservoir. We characterize the which-path information about the resonator's state with the distinguishability. The experimental results reveal that the amount of spontaneously erasable which-path information progressively decreases with the increase in the number of the reservoir's degrees of freedom. When the reservoir contains sufficiently many oscillators, the which-path information acquired by the reservoir cannot be erased by the system-reservoir interaction at later times, which makes the decoherence irreversible. We further characterize the non-Markovianity of such a controllable open mesoscopic system.
\end{abstract}

\maketitle

The superposition principle lies at the heart of quantum mechanics and discriminates between the microscopic and macroscopic worlds \cite{Zurek1981PRD,Zurek1982PRD,Zurek1991PT,Zurek2003RMP}. It allows a quantum system to be simultaneously in two states corresponding to different values of an observable, whose quantum interference gives rise to new effects that cannot be interpreted in classical terms. The validity of this principle has passed a large number of experimental tests and lays the foundation for quantum technology, ranging from quantum information transmission and processing \cite{NielsenChuang2000} to quantum metrology \cite{Degen2017RMP,McCormick2019Nature,Wang2019NatCommun}. Although quantum superpositions are ubiquitous in the microscopic world, they are difficult to observe in the macroscopic world, as illustrated by the famous Schr\"{o}dinger cat paradox \cite{Schrodinger1935}.
  Nowadays, it is well known that environment-induced decoherence is responsible for the disappearance of macroscopic superpositions \cite{Zurek1981PRD,Zurek1982PRD,Zurek1991PT,Zurek2003RMP}. Any quantum system would unavoidably couple to its surrounding environment, which corresponds to a reservoir that involves many degrees of freedom. A zero-temperature reservoir can be thought of as an ensemble of electromagnetic modes, each of which is initially in its ground state. The coupling between the system and reservoir results in their entanglement. In terms of complementarity, this entanglement enables the environment to continuously gather the which-path information about the system \cite{Brune1996PRL,Scully1991Nature,Englert1996PRL,Buks1998Nature,Durr1998PRL,Bertet2001Nature,Huang2023npjQI,Hornberger2003PRL,Hackermuller2004Nature}, destroying the quantum coherence between the superimposed states (paths), thereby transforming their quantum superposition into a classical mixture. This decoherence dynamics occurs at a rate linearly scaling with the size of the quantum superposition. Quantum Darwinism expands on the decoherence theory, stating that the environment redundantly encodes information about the quantum system, which is responsible for the emergence of classical reality \cite{Zurek2009NatPhys,Touil2024Quantum,Zhu2025SciAdv}.

The quantum coherence for a macroscopic meter is lost too fast to observe. However, a mesoscopic superposition of states, which is distinguished by some classical property, offers the possibility for exploring the decoherence process and the quantum-to-classical transition \cite{Haroche2013RMP,Wineland2013RMP}. Such mesoscopic cat states are usually defined as superpositions of coherent states with distinct amplitudes or phases of a bosonic mode. In addition to fundamental interest, such superposition states are useful for fault-tolerant quantum computation \cite{Ofek2016Nature,Grimm2020Nature} and quantum metrology \cite{Toscano2006PRA,Penasa2016PRA,Vlastakis2013Science,Tan2025PRXQ,Zheng2025PRA}. Over the past three decades, enduring efforts have been devoted to experimental investigations of irreversible decoherence processes of these mesoscopic superpositions in different systems, including photonic fields \cite{Brune1996PRL,Haroche2013RMP,Deleglise2008Nature}, motional modes of trapped ions \cite{Myatt2000Nature,Turchette2000PRA,Myatt2000JModOpt}, and mechanical oscillators \cite{Bild2023Science}. The decoherence dynamics of an open system depends upon the spectral structure of the associated reservoir. In particular, the decoherence of a photonic cat state stored in a cavity becomes completely reversible when this cavity is coupled to a second cavity, which serves as a non-Markovian ``single-mode reservoir" \cite{Raimond1997PRL}. Under this system-reservoir interaction, the mesoscopic quantum coherence would exhibit periodical collapses and revivals due to the fact that the which-path information encoded in the reservoir mode can be erased by interaction at later times. When more and more ``reservoir modes" are involved, the standard irreversible decoherence model is recovered. The reversibility of decoherence is closely related to the non-Markovianity of the reservoir \cite{Wolf2008PRL,Breuer2009PRL,Rivas2010PRL,Rivas2014RPP,Breuer2016RMP,Fanchini2014PRL,Haase2018PRL,Peng2018SciBull}, which characterizes the reservoir's memory effects. The transition from the non-Markovian to Markovian regime has been observed in different experiments \cite{Liu2011NatPhys,Wang2018PRB,Li2022PRL,Gaikwad2024PRL,Han2024APL}, but they all focused on pseudospin dynamics in controlled reservoirs.

Here we report an experimental demonstration of the reversible-to-irreversible transition of decoherence for photonic cat states. The experiment is performed with a circuit quantum electrodynamics (QED) architecture, where a bus resonator, initially prepared in a quantum superposition of being empty and filled with a coherent field, is controllably coupled to an engineered reservoir composed of Josephson-junction-based nonlinear oscillators, each corresponding to a qubit. The reservoir acquires the which-path information of the system by energy exchange, which is reversible for the single-qubit case. Such a process becomes more chaotic with the increase in the number of the reservoir qubits, since different qubits exchange energy with the field mode with different rates. Consequently, the which-path information acquired by the reservoir cannot be spontaneously erased by interaction at later times, when the reservoir contains sufficiently many qubits. We extract the which-path information encoded in the reservoir from the measured output density matrices of the qubits. With the increase in the number of reservoir qubits, the amount of spontaneously erasable which-path information decreases, and the revivable quantum coherence of the photonic cat state reduces. The experimental results illustrate the transition from non-Markovian to Markovian dynamics of mesoscopic cat states.

The system under investigation is a bosonic field mode, which is coupled to a reservoir involving $N$ nonlinear oscillators. Each of these reservoir oscillators is confined within its lowest two levels, and thus corresponds to a qubit. In the framework rotating at the system frequency, the system-reservoir interaction is described by the Hamiltonian ($\hbar=1$)
\begin{equation}
H=\sum_{k=1}^{N}\left[ \frac{\lambda_{k}}{2}\left(a\left|e\right\rangle_{k}\left\langle g\right|+a^{\dagger}\left|g\right\rangle_{k}\left\langle e\right|\right)\right],
\end{equation}
where $a^{\dagger}$ and $a$ represent the creation and annihilation operators for the system field mode, and $\left|e\right\rangle_{k}$ and $\left|g\right\rangle_{k}$ denote the upper and lower levels for the $k$th qubit, which is resonantly coupled to the system with strength $\lambda_{k}/2$. We first consider the case that the reservoir only involves a single qubit and the field mode is initially in a coherent state $\left|\alpha\right\rangle$, which represents a quantum state with a close classical analog and can be expanded in terms of photon-number states,
\begin{equation}
\left|\alpha\right\rangle=e^{-\left|\alpha\right|^{2}/2}\sum_{n=0}^{\infty}\frac{\alpha^{n}}{\sqrt{n!}}\left|n\right\rangle.
\end{equation}
Suppose that the average photon number, $\left\langle n\right\rangle=\left|\alpha\right|^{2}$, is much larger than the width of the Poissonian photon-number distribution $\delta n=\left|\alpha\right|$. The qubit, initially in the ground state $\left|g\right\rangle_{1}$, undergoes Rabi oscillations with the period $T=2\pi/\Omega_{1}$, where $\Omega_{1}=\sqrt{\left\langle n\right\rangle}\lambda_{1}$ is the Rabi frequency. During the interaction, the qubit produces a backaction on the coherent field by splitting it into two components rotating at the angular velocities $\pm\omega_{1}=\pm\lambda_{1}^{2}/(4\Omega_{1})$ \cite{Auffeves2003PRL,Zheng2013QIC}. When the time $t$ is much smaller than $\pi/\omega_{1}$, this backaction can be neglected, so that the field mode approximately remains in the quasiclassical state $\left|\alpha\right\rangle$.

We now turn to the decoherence process of an amplitude photonic cat state during interaction with a single-qubit reservoir. Such a cat state is formed by the superposition of the vacuum state and coherent state $\left|\alpha\right\rangle$ with $\left|\alpha\right|^{2}\gg 1$,
\begin{equation}
\left|{\cal C}\right\rangle\simeq(\left|0\right\rangle+\left|\alpha\right\rangle)/\sqrt{2}.
\end{equation}
As the qubit is initially in its ground state, it is decoupled from the field's vacuum component $\left|0\right\rangle$. Under the above condition, the conditional Rabi dynamics evolves the qubit and the resonator into an entangled state, given by $[\left|0\right\rangle\left|g\right\rangle_{1}+\left|\alpha\right\rangle(c_{1}^{g}\left|g\right\rangle_{1}+c_{1}^{e}\left|e\right\rangle_{1})]/\sqrt{2}$, where $c_{1}^{g}=\cos\theta$ and $c_{1}^{e}=-i\sin\theta$ with $\theta=\Omega_{1}t/2$. This entanglement partially encodes the information about the field amplitude in the qubit state, which results in the degradation of quantum coherence between $\left|0\right\rangle$ and $\left|\alpha\right\rangle$. When the reservoir qubit is traced out, the system field is left in a mixed state, described by the density operator
\begin{equation}
\rho_{s}\simeq\left[|0\rangle\langle0|+|\alpha\rangle\langle\alpha|+c_{1}^{g}\left(|0\rangle\langle\alpha|+|\alpha\rangle\langle0|\right)\right]/2.
\end{equation}
Consequently, the quantum coherence is reduced by a factor of $\left|c_{1}^{g}\right|$. As the which-path information encoded in the qubit can be erased by the field-qubit interaction at later times, the decoherence process is reversible. Therefore, the qubit acts as a non-Markovian reservoir for the mesoscopic field.

When the field mode is coupled to a bath involving $N$ qubits, each initially in its ground state, the total system-reservoir state approximately evolves as
\begin{equation}
\left|\psi(t)\right\rangle\simeq\left(\left|0\right\rangle\left|\phi^{_{\left|0\right\rangle}}\right\rangle+\left|\alpha\right\rangle\left|\phi^{_{\left|\alpha\right\rangle}}\right\rangle\right)/\sqrt{2},
\end{equation}
where $\left\vert \phi ^{_{\left\vert 0\right\rangle }}\right\rangle = \bigotimes_{k=1}^N
\left\vert g\right\rangle _{k}$ and $\left\vert \phi
^{_{\left\vert \alpha \right\rangle }}\right\rangle = \bigotimes_{k=1}^N (c_{k}^{g}\left\vert
g\right\rangle _{k}+c_{k}^{e}\left\vert e\right\rangle _{k})$.
The approximation is valid only when the coherent state component $\left|\alpha\right\rangle$ is not seriously affected by interaction with the reservoir qubits. When all degrees of freedom of the reservoir are traced out, the system's density operator reduces to
\begin{equation}
\rho_{s}\simeq\frac{1}{2}\left[|0\rangle\langle0|+|\alpha\rangle\langle\alpha|+\left(\prod_{k=1}^{N}c_{k}^{g}|\alpha\rangle\langle0|+{\rm H.c}\right)\right].
\end{equation}
If the reservoir contains many qubits that have different Rabi frequencies, the phases of their Rabi oscillations cannot be resynchronized so that the which-path information encoded in the collective state of the qubits cannot be spontaneously erased by the system-reservoir interaction at later times, leading to an irreversible decoherence process.

\begin{figure}[t]
\centering
\includegraphics[width=1\linewidth]{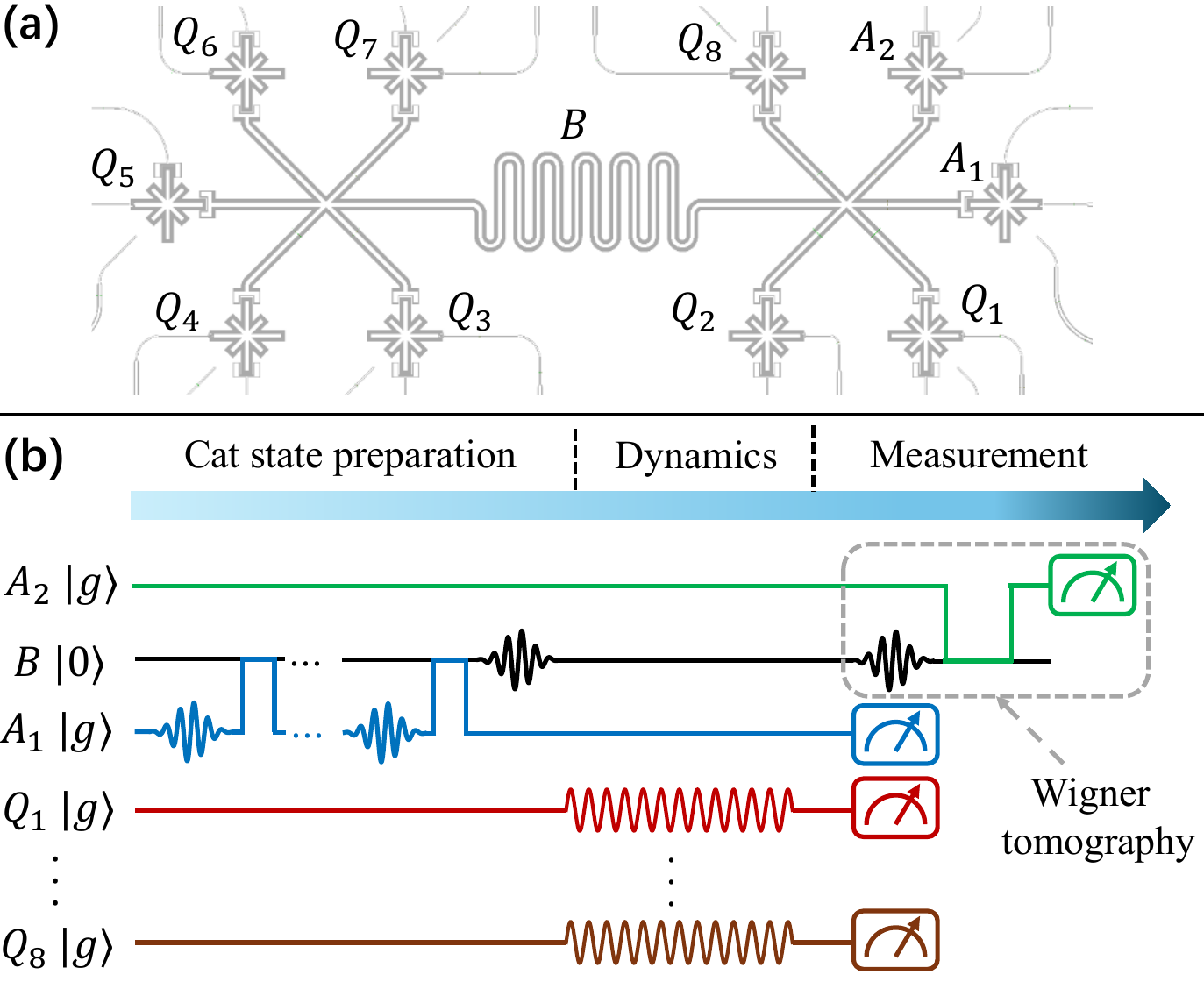}
\caption{Overview of the device and experimental pulse sequence. (a) Sketch of the device. The device contains a frequency-fixed bus resonator ($B$) connected to 10 frequency-tunable Josephson-junction-based nonlinear oscillators, each of which is confined to the lowest two levels and behaves as a qubit. Two of these qubits, denoted as $A_{1}$ and $A_{2}$, are used as ancillae to prepare the photonic cat state of the photonic mode stored in $B$ and to probe its quantum coherence, respectively. The other qubits, labeled from $Q_{1}$ to $Q_{8}$, serve as the reservoir oscillators that can be selectively coupled to $B$. (b) Experimental pulse sequence.}
\label{fig1}
\end{figure}

The model is realized in a circuit QED architecture, where 10 frequency-tunable Josephson-junction-based nonlinear oscillators are capacitively coupled to a bus resonator ($B$) with a fixed frequency $\omega_{s}/(2\pi)=5.796$ GHz, as sketched in Fig.~\ref{fig1}(a) \cite{Song2017PRL,Zheng2025PRL_SSB}. The photonic mode stored in $B$ corresponds to the system field mode. Two of these nonlinear oscillators, denoted as $A_{1}$ and $A_{2}$, are respectively used as ancillae for preparing the cat state and for detecting the quantum coherence of the system field mode stored in $B$, while the others serve as the reservoir oscillators, represented by $Q_{j}$ ($j=1$ to $8$). The on-resonance coupling strengths between these nonlinear oscillators and $B$ range from $2\pi\times 13$ to $2\pi\times 21$~MHz. Each nonlinear oscillator has an anharmonicity $\sim 2\pi \times 250$ MHz, which enables it to behave well as a qubit. Before the experiment, each of the qubits stays at its idle frequency, where it is effectively decoupled from the system resonator as the detuning is much larger than the corresponding coupling strength. The system parameters are detailed in Sec. S1 of Supplemental Material \cite{supp}. The experiment starts with the preparation of the cat state, following the step-by-step technique proposed in the context of cavity QED \cite{LawEberly1996PRL} and demonstrated in ion trap \cite{BenKish2003PRL} and circuit QED experiments \cite{Hofheinz2009Nature,Zheng2015PRL_DelayedChoice}. To shorten the pulse sequence, we first drive the resonator from the initial ground state to the phase cat state $(\left\vert -\alpha /2\right\rangle+\left\vert \alpha /2\right\rangle )/\sqrt{2}$ using $A_{1}$, initially in its ground state, which coherently pumps photons from an external microwave source into $B$ one by one (see Sec. S2 in Supplemental Material \cite{supp} for details), thanks to the tunability of the frequency of $A_{1}$, enabled by an ac flux. The size of the cat state is characterized by the parameter $\alpha$, which has a value of $3.3$ in our experiment. The Uhlmann fidelities of the experimentally prepared phase and amplitude cats are $0.929$ and $0.913$, respectively.

 Theoretically, $A_1$ is left in $|g\rangle_1$ after the photonic cat state preparation and can be used for subsequent state tomography of $B$. However, because of experimental imperfections, $A_1$ retains a small $|e\rangle_1$-state population, which would reduce the readout fidelity of $B$. To avoid this problem, we use $A_2$ to read out the state of $B$.

\begin{figure}[t]
\centering
  \includegraphics[width=1\linewidth]{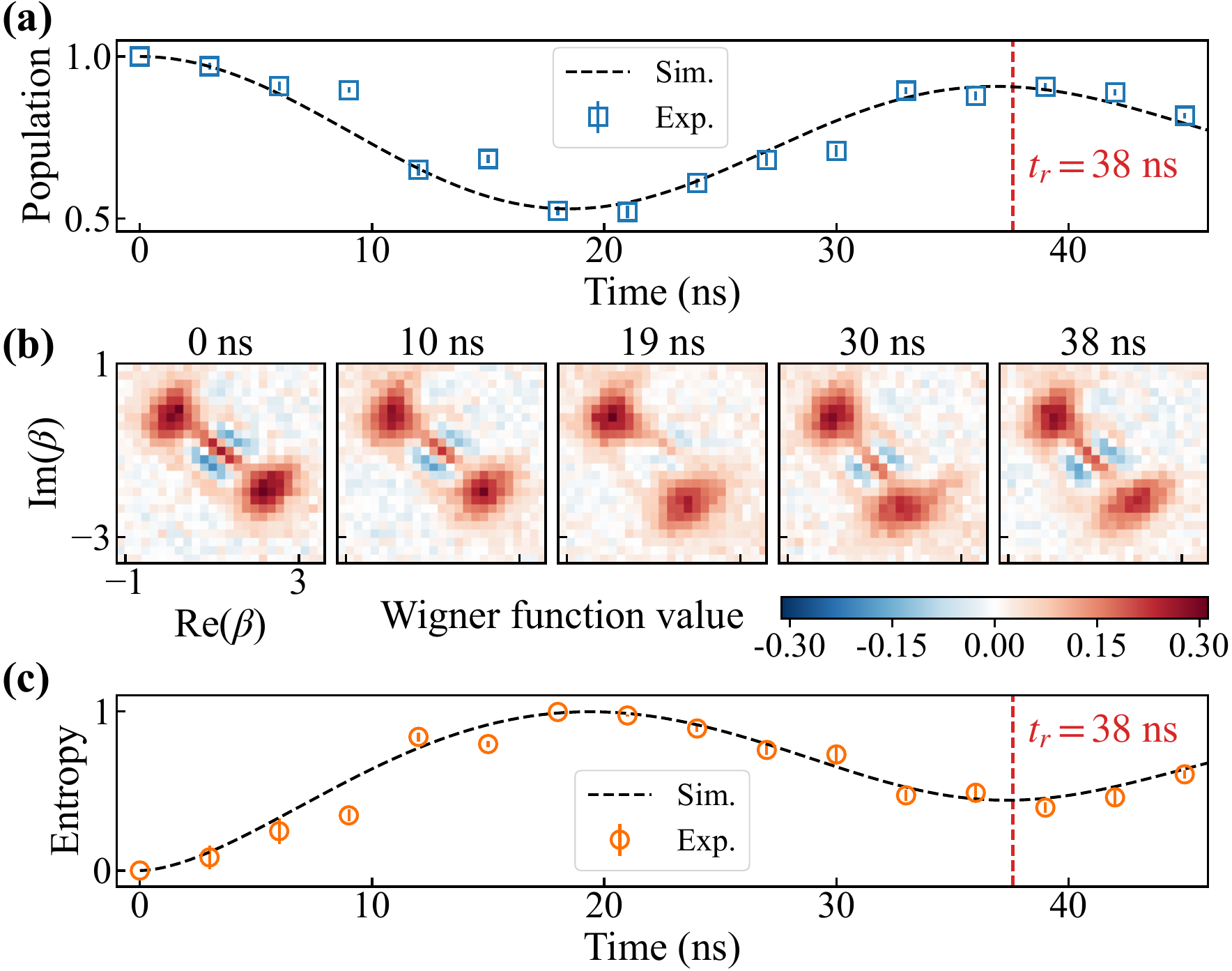}
 \caption{Observation of the reversible decoherence of the cat state induced by a single reservoir qubit ($Q_{1}$). (a) The $\left\vert g\right\rangle$-state population of $Q_{1}$ measured for different interaction times $t$. (b) Wigner functions of the mesoscopic field for different $B$-$Q_{1}$ interaction times. (c) Evolution of the von Neumann entropy of $Q_{1}$. Before the interaction, $B$ is prepared in the amplitude cat state of Eq. (3) with $\alpha=3.3$ and $Q_{1}$ is initialized to its ground state $\left\vert g\right\rangle_{1}$. The dashed lines in (a) and (c) denote the numerical results. Error bars indicate the standard deviation and the vertical red dashed line marks the revival time $t_r = 2\pi/\Omega_1 \approx 38~\mathrm{ns}$, which corresponds to the period of the Rabi oscillations.}
\label{fig2}
\end{figure}

We first consider the decoherence behavior of the photonic cat state coupled to a single reservoir qubit ($Q_{1}$). Figure~\ref{fig1}(b) shows the pulse sequence for realizing this reversible decoherence process and probing the quantum coherence after this process. We first transform the phase cat state into the desired amplitude cat state by the phase-space displacement $D(\alpha/2)$. The qubit $Q_1$ is first parked at the static frequency $\omega_s-\nu_1$. An ac-flux tone at frequency $\nu_1$ then brings it to resonance with $B$ at the first upper Floquet sideband of the modulation, with effective coupling strength $\lambda_{1}/2=J_{1}(\mu_{1})\xi_{1}$ \cite{Li2018PRApp,Wang2019NatPhys,Zheng2023PRL_PhaseTransition}, where $\xi_{1}$ is the on-resonance $B$-$Q_{1}$ photon swapping rate and $J_{1}(\mu_{1})$ represents the first-order Bessel function of the first kind, with $\mu_{1}=\varepsilon_{1}/\nu_{1}$. The resulting $B$-$Q_{1}$ interaction is described by the effective Hamiltonian of Eq. (1) with $N=1$. In our experiment, $\lambda_{1}$ is set to $2\pi\times 8.1$ ~MHz. After a preset interaction time, the modulation is switched off. The evolution of the measured $\left\vert g\right\rangle$-state population of $Q_{1}$, $P_{g}$, is displayed in Fig.~\ref{fig2}(a). The observed Rabi oscillation has a period of about $t_r=38$~ns, which is in good agreement with the theoretical value, given by $T=2\pi/\Omega_1=37.4$~ns, where $\Omega_1\simeq|\alpha|\lambda_1=2\pi\times26.7$~MHz.

\begin{figure}[htbp]
\centering
  \includegraphics[width=1\linewidth]{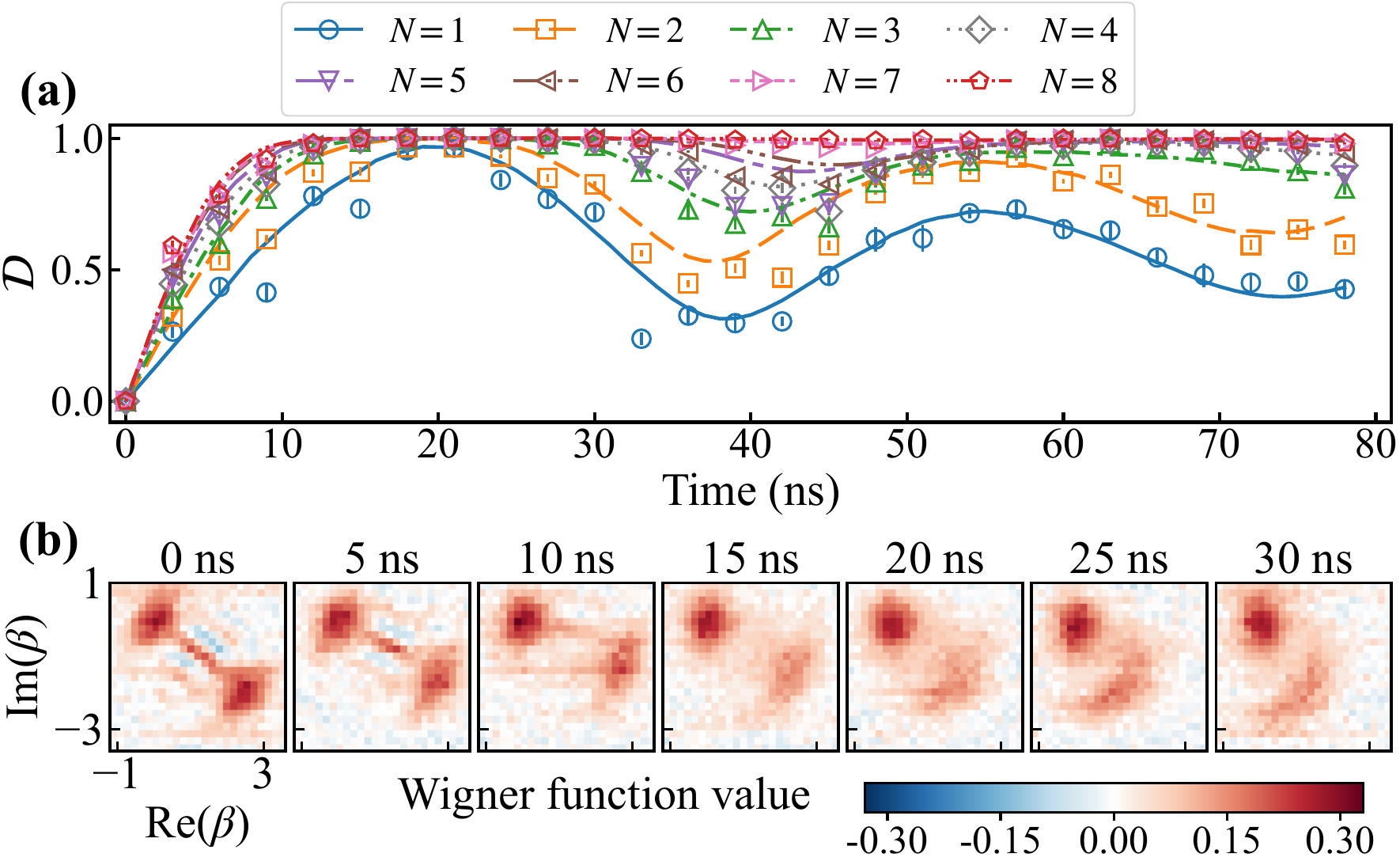}
 \caption{Progressive transition from reversible to irreversible decoherence. (a) Evolutions of distinguishabilities for different values of $N$. $N$ ($N=1$ to $8$) reservoir oscillators are selectively coupled to the system, prepared in the amplitude cat state. The effective coupling strength associated with each qubit is controlled by the parameters of the modulating pulses applied to mediate the corresponding sideband interactions. Error bars indicate the standard deviation. The markers represent the experimental data, and the curves represent the corresponding calculations for the experimentally prepared cat state. (b) Wigner functions of the mesoscopic field for $N=8$, measured for different interaction times.}
\label{fig3}
\end{figure}

To illustrate the resulting decoherence process of the mesoscopic field, we perform Wigner tomography \cite{Deleglise2008Nature,Hofheinz2009Nature} (see Sec. S5B in Supplemental Material \cite{supp} for details). The Wigner function of the field describes the phase-space quasiprobability distribution. The value of this function at position $\beta$, denoted as $W(\beta)$, is directly related to the expectation value of the photon number parity after the phase-space displacement $D(-\beta)=e^{-\beta a^{\dagger}+\beta^{\ast}a}$. The photon-number distribution of the displaced field state is inferred by the Rabi signals of $A_{2}$ resonantly coupled to the resonator. The reliability of the parity reconstruction at large phase-space displacements is quantitatively assessed in Sec. S5B of Supplemental Material. The Wigner functions, reconstructed for different $B$-$Q_{1}$ interaction times, are displayed in Fig.~\ref{fig2}(b). The quantum coherence of the initial cat state is manifested by the interference fringes with alternating positive and negative values between the two Gaussian peaks, each corresponding to one quasiclassical state component. At time $t=t_r$, the qubit nearly makes a full cycle of Rabi oscillations when the field is in the coherent state $\left\vert \alpha\right\rangle$, and consequently, the quantum coherence is nearly revived. In terms of complementarity, the qubit can be considered as a which-path detector in phase space. The path information is encoded in the qubit's state and then erased by the unitary dynamics at later times leading to disappearance and revival of the phase-space interference.

The $B$-$Q_{1}$ entanglement can be characterized by the von Neumann entropy of the output state of $Q_{1}$, defined as ${\cal S}=-{\rm Tr}(\rho_{q}\log_{2}\rho_{q})$, where $\rho_{q}$ denotes the reduced density matrix of $Q_{1}$. For each interaction time, we measure the three Bloch-vector components of $Q_1$, reconstruct its $2\times2$ density matrix, and evaluate the entropy from its eigenvalues. Figure~\ref{fig2}(c) presents the entropy measured for different $B$-$Q_{1}$ interaction times. As expected, the entropy also exhibits an oscillatory pattern. During the first stage, ${\cal S}$ quickly increases, reaching the maximum at time $t=19$ ns when $B$ and $Q_{1}$ evolve to a maximally entangled state. Then, ${\cal S}$ drops and evolves to a local minimum value at $t=t_r$. The residual $\left\vert e\right\rangle_{1}$-state population and entropy observed at $t=t_r$ are mainly due to the spread of the phases for the quantum Rabi oscillations associated with the dominant Fock-state components in $\left\vert \alpha \right\rangle$, which is about $\Omega_{1}T\delta n/(2\left\langle n\right\rangle)=\pi/\alpha$ for $\left\langle n\right\rangle\gg 1$ \cite{Scully1997QuantumOptics}. With the increase in $\alpha$, this spread is narrowed, so that the revival becomes more perfect (see Sec. S9 of Supplemental Material \cite{supp} for details).

The transition from reversible to irreversible decoherence is illustrated by gradually increasing the number of qubits ($N$) that are effectively coupled to the bus resonator. To make the coherent state $\left\vert \alpha \right\rangle$ remain quasiclassical during the interaction, we suitably set the effective strengths of the sideband interactions so that the total number of photons transferred from $\left\vert \alpha \right\rangle$ to the qubits is much smaller than $\left\vert \alpha \right\vert^{2}$. The frequencies and amplitudes of the modulations, used to mediate sideband interactions between the bus resonator and reservoir oscillators, are detailed in Sec. S4 of Supplemental Material \cite{supp}.

  When more qubits are involved, there is no one-to-one correspondence between the entropies of individual qubits and the amount of which-path information encoded in the reservoir. In this case, it is convenient to quantify the which-path information stored in the reservoir with the distinguishability ${\cal D}$ \cite{Englert1996PRL}, defined as
\begin{equation}
{\cal D}=\frac{1}{2}{\rm Tr}\left|\rho^{\left|0\right\rangle}-\rho^{\left|\alpha\right\rangle}\right|,
\end{equation}
where $\rho^{\left|0\right\rangle}$ and $\rho^{\left|\alpha\right\rangle}$, respectively, denote the reduced density operator for the reservoir qubits associated with the field components $\left\vert 0\right\rangle$ and $\left\vert \alpha\right\rangle$ after the system-reservoir interaction. The value of ${\cal D}$ ranges from 0 to 1, depending upon the amount of which-path information that can be extracted from the reservoir in principle. For $\rho^{\left|0\right\rangle}=\rho^{\left|\alpha\right\rangle}$, ${\cal D}=0$. In this case, the reservoir contains no information about whether the field mode is in $\left\vert 0\right\rangle$ or $\left\vert \alpha\right\rangle$, so that the quantum coherence remains intact. When $\rho^{\left\vert 0\right\rangle}$ and $\rho^{\left\vert \alpha\right\rangle}$ are orthogonal, ${\cal D}=1$, which corresponds to the case that the reservoir contains unambiguous information about the amplitude of the field mode, destroying the system's quantum coherence. As each reservoir qubit remains in the ground state $\left\vert g\right\rangle_{k}$ for the vacuum field state $\left\vert 0\right\rangle$, we have $\rho ^{\left\vert 0\right\rangle }= \bigotimes_{k=1}^N \left\vert g\right\rangle _{k}\left\langle g\right\vert$. Suppose that the reduced density operator of the $k$th reservoir qubit after the interaction is $\rho_{k}^{\left\vert \alpha\right\rangle}$ when the field is initially in $\left\vert \alpha\right\rangle$. This implies that its reduced density matrix ($\rho_{k}$) after the interaction is the equal mixture of $\left\vert g\right\rangle_{k}\left\langle g\right\vert$ and $\rho_{k}^{\left\vert \alpha\right\rangle}$ when the field is initially in the cat state. Therefore, $\rho_{k}^{\left\vert \alpha\right\rangle}$ is estimated from the measured $\rho_{k}$: $\rho_{k}^{\left\vert \alpha\right\rangle}=2\rho_{k}-\left\vert g\right\rangle_{k}\left\langle g\right\vert$. $\rho^{\left\vert \alpha\right\rangle}$ is then approximated by $\rho^{\left\vert \alpha \right\rangle }= \bigotimes_{k=1}^N \rho _{k}^{\left\vert \alpha \right\rangle }$. The detailed derivation of this approximation is given in Sec. S6B of Supplemental Material \cite{supp}.

Figure~\ref{fig3}(a) shows the evolutions of the distinguishabilities for different values of $N$, inferred from the individually measured output states of the reservoir oscillators. The curves represent the results calculated with the experimentally prepared cat state. As expected, for $N=1$, ${\cal D}$ exhibits an oscillatory pattern. When $N$ is increased, the oscillatory pattern is gradually washed out. For $N=8$, ${\cal D}$ almost remains unchanged after it reaches the maximum at the time $t=15$ ns. These results can be interpreted as below. The transition $\left\vert g\right\rangle\rightarrow\left\vert e\right\rangle$ for a single reservoir qubit associated with $\left\vert \alpha \right\rangle$ is sufficient to determine which quasiclassical component the field mode is in. As the Rabi oscillations of different reservoir qubits are nonsynchronous, the probability of all qubits returning to their ground states decreases when $N$ is increased. Consequently, with the increase in $N$, the which-path information acquisition process becomes more and more irreversible. We also perform an additional experiment by separately preparing $|0\rangle$ and $|\alpha\rangle$ for $N=1$--$8$. The results are presented in Sec. S6D and Fig. S11 of Supplemental Material \cite{supp}. The distinguishabilities obtained from these separate experiments agree well with the results shown in Fig.~\ref{fig3}(a).

To confirm that the distinguishability is closely related to the loss of quantum coherence between the two quasiclassical components $\left\vert 0\right\rangle$ and $\left\vert \alpha \right\rangle$ of the field mode, we observe the evolution of the output field state for $N=8$. The Wigner functions of the field mode, measured for different interaction times with the reservoir, are displayed in Fig.~\ref{fig3}(b). As expected, the interference fringes between the two Gaussian peaks are gradually washed out during the interaction, without revival. This coincides with the evolution of ${\cal D}$, which also exhibits irreversible behavior in this case. The deformation of the coherent-state component during the evolution is caused by the self-Kerr effect produced by the interaction with the qubits, which results in phase spreading \cite{Auffeves2003PRL}. The observed irreversible decoherence phenomenon originates from a unitary system-environment interaction, which is responsible for the emergence of classicality, as described by quantum Darwinism \cite{Zurek2009NatPhys}. We further experimentally verify that the non-Markovianity of the artificial reservoir monotonically decreases with the increase of $N$, as detailed in Sec. S10 of Supplemental Material \cite{supp}.

In conclusion, we have investigated the decoherence process of a photonic cat state coupled to an artificial reservoir formed by nonlinear oscillators. By coupling the photonic mode to more and more reservoir oscillators, we observe the progressive reversible-to-irreversible transition of the decoherence process of the mesoscopic field. Our results reveal that the amount of lost quantum coherence between the two superimposed quasiclassical components of the system depends upon the amount of which-path information encoded in the reservoir. When the number of the reservoir oscillators is sufficiently large, this information, once acquired by the reservoir, cannot be spontaneously erased during the interaction at later times due to the nonsynchronization of the quantum oscillations associated with distinct reservoir oscillators. Consequently, the observed phase-space quantum interference effect irreversibly decays. The techniques developed in the experiment may find applications in quantum control of complex circuit QED systems involving many qubits and bosonic modes, which represent a promising platform for realizing quantum technologies.

\nocite{Xu2020PRL}

\begin{acknowledgments}
This work was supported by the National Natural Science Foundation of China (Grants No. 12274080, No. 12474356, No. 12475015, and No. 62471143), the Key Program of the National Natural Science Foundation of Fujian Province (Grant No. 2024J02008), and the Innovation Program for Quantum Science and Technology (Grant No. 2021ZD0300200).
\end{acknowledgments}

\noindent\textit{Data availability}---The data that support the findings of this article are available from the corresponding author upon reasonable request.

%

\end{document}


\setcounter{secnumdepth}{3}

\title{Supplemental Material for ``Observation of the Transition from Reversible to Irreversible Decoherence of Mesoscopic Quantum Superpositions''}

\author{Ri-Hua Zheng}
\author{Jia-Hao L\"{u}}
\author{Fan Wu}
\author{Yan Xia}\email{xia-208@163.com}
\affiliation{Fujian Key Laboratory of Quantum Information and Quantum Optics, College of Physics and Information Engineering, Fuzhou University, Fuzhou, Fujian, 350108, China}

\author{\\ Li-Hua Lin}
\author{ Zhen-Biao Yang}\email{zbyang@fzu.edu.cn}
\author{Shi-Biao Zheng}\email{t96034@fzu.edu.cn}
\affiliation{Fujian Key Laboratory of Quantum Information and Quantum Optics, College of Physics and Information Engineering, Fuzhou University, Fuzhou, Fujian, 350108, China}
\affiliation{Hefei National Laboratory, Hefei 230088, China}

\maketitle

\tableofcontents

\newpage

\section{Qubit and Resonator Parameters}\label{QR_Parameters}

\begin{table}[t]
\begin{center}
\setlength{\tabcolsep}{6pt}
\begin{tabular}{lcccccccccccc}
\hline\hline
 Parameters & $Q_1$ & $Q_2$ & $Q_3$ & $Q_4$ & $Q_5$ & $Q_6$ & $Q_7$ & $Q_8$ & $A_1$ & $A_2$ \\
\hline
$\omega_{i}/(2\pi)$ (GHz) & 5.30 & 5.25 & 5.15 & 5.35 & 4.95 & 5.05 & 5.00 & 5.12 & 5.20 & 5.10 \\
$F_{g}$ & 0.95 & 0.97 & 0.95 & 0.93 & 0.92 & 0.95 & 0.94 & 0.93 & 0.95 & 0.94 \\
$F_{e}$ & 0.89 & 0.88 & 0.86 & 0.85 & 0.79 & 0.87 & 0.87 & 0.88 & 0.81 & 0.87 \\
$\omega_{r}/(2\pi)$ (GHz) & 6.695 & 6.697 & 6.638 & 6.617 & 6.619 & 6.544 & 6.511 & 6.892 & 6.811 & 6.792 \\
$\omega_{h}/(2\pi)$ (GHz) & 5.775 & 5.777 & 5.714 & 5.723 & 5.728 & 5.746 & 5.748 & 5.786 & 5.923 & 5.822 \\
$T_{1}$ ($\mu$s) & 14.9 & 29.8 & 15.6 & 14.4 & 13.1 & 24.7 & 26.8 & 15.5 & 25.5 & 14.0 \\
$T_{2}$ ($\mu$s) & 3.0 & 2.4 & 2.6 & 3.2 & 1.6 & 2.0 & 2.1 & 2.2 & 2.5 & 2.0 \\
$T_{2}^{\rm SE}$ ($\mu$s) & 11.7 & 10.1 & 10.0 & 10.3 & 8.5 & 10.2 & 8.6 & 8.6 & 9.3 & 8.0 \\
$\xi$ (2$\pi$ MHz) & 19.6 & 19.9 & 16.0 & 20.2 & 19.2 & 20.5 & 12.9 & 16.3 & 19.8 & 19.8 \\
\hline\hline
\end{tabular}
\end{center}
\caption{\textbf{Performance of superconducting qubits.}
The table lists ten qubits ($Q_1$–$Q_8$ and $A_1$–$A_2$).
The entry $\omega_i/(2\pi)$ denotes the idle frequency used as the reference for state preparation and standby operation.
Frequency tuning is implemented through the $Z$ control line by applying direct-current offsets together with square-wave bias pulses, which shift the qubit frequency; the highest reached frequency is $\omega_h/(2\pi)$.
Readout fidelities are $F_g$ and $F_e$.
Coherence parameters $T_1$, $T_2$, and $T_2^{\rm SE}$ are measured at each qubit’s idle frequency, with $T_1$ representing energy relaxation, $T_2$ obtained from Ramsey dephasing (Gaussian fit), and $T_2^{\rm SE}$ from spin echo (exponential fit).
Each qubit is read out through a dedicated resonator of frequency $\omega_r/(2\pi)$ and couples to the common bus (storage) resonator with strength $\xi$.}
\label{tab:qr_all}
\end{table}

The device comprises ten superconducting qubits, labeled $Q_1$–$Q_8$ and $A_1$–$A_2$, whose performance metrics are summarized in Table~\ref{tab:qr_all}.
Qubit operation is referenced to the idle point $\omega_i/(2\pi)$: this is the frequency where qubit states are prepared and where a qubit resides whenever it is not involved in dynamical protocols.
Frequency excursions are realized via the $Z$ control line, which allows fast tuning of the qubit frequency; the highest accessible point reached in this way is denoted $\omega_h/(2\pi)$.
Coherence metrics are characterized at the idle point, with $T_1$ quantifying energy relaxation, $T_2$ obtained from Ramsey interferometry, and $T_2^{\rm SE}$ extracted with spin–echo.
State discrimination is evaluated by $F_g$ and $F_e$.
Each qubit is dispersively coupled to its dedicated readout resonator of frequency $\omega_r/(2\pi)$, and couples to the common bus resonator (serving as the storage mode) with strength $\xi$.

For the bus (storage) resonator, the fixed frequency is $\omega_{s}/(2\pi)=5.796~\mathrm{GHz}$, with an energy-relaxation time of $T_{1,s}=3.0~\mu\mathrm{s}$ and a dephasing time of $T_{2,s}=3.8~\mu\mathrm{s}$.
Both $T_{1,s}$ and $T_{2,s}$ are extracted by resonantly tuning a qubit into the cavity and swapping the cavity excitations to the qubit.
The relaxation of the mapped excitation yields $T_{1,s}$, while Ramsey measurement on the qubit after the swap provides $T_{2,s}$.

\section{Preparation of Initial Cat States in the Resonator}

\subsection{Phase-cat preparation via sequential photon-number swaps}

We target the even cat
\begin{equation}
|C_{+}(\alpha/2)\rangle=\mathcal N_{+}\big(|\alpha/2\rangle+|-\alpha/2\rangle\big),\qquad
\mathcal N_{+}=\Big[2\big(1+e^{-|\alpha|^2/2}\big)\Big]^{-1/2},
\end{equation}
whose Fock decomposition contains only even photon numbers,
\begin{equation}
|C_{+}(\alpha/2)\rangle=\sum_{m=0}^{\infty} c_{2m}\,|2m\rangle,\qquad
c_{2m}=\mathcal N_{+}\,\frac{2(\alpha/2)^{2m}}{\sqrt{(2m)!}}\,e^{-|\alpha|^2/8}.
\end{equation}
Since only even components are present and the coefficients \(c_{2m}\) decrease rapidly with \(m\), we work in a truncated resonator subspace. For the cat size used here \((\alpha/2=1.65)\), more than \(98.9\%\) of the probability lies in \(n\le 6\). We therefore fix an operational cutoff \(N^\star=6\) and design the time-reversed sequence to explicitly empty the \(n\)-excitation manifolds \(\mathrm{span}\{|g,n\rangle,|e,n-1\rangle\}\) for \(n=6,5,\dots,1\); components with \(n>6\) are left untouched and contribute only a negligible tail to the truncated-subspace overlap.

On each \(n\)-excitation manifold, a resonant swap \(S_n\) (under the Jaynes–Cummings interaction) of duration \(t_n\) performs
\begin{equation}
\begin{pmatrix}a'_{g,n}\\ a'_{e,n-1}\end{pmatrix}
=
\begin{pmatrix}
\cos\theta_n & -i\sin\theta_n\\
-i\sin\theta_n & \cos\theta_n
\end{pmatrix}
\begin{pmatrix}a_{g,n}\\ a_{e,n-1}\end{pmatrix},
\qquad
\theta_n=\sqrt{n}\,\xi\,t_n,
\label{eq:Sn-matrix}
\end{equation}
where \(\xi\) denotes the qubit–resonator coupling strength. A single-qubit rotation \(Q_n\) within \(\{|g,n-1\rangle,|e,n-1\rangle\}\) is taken as
\begin{equation}
Q_n = X_{\pi}
=\exp\!\big(-i\tfrac{\pi}{2}\sigma_x\big),
\end{equation}
and no additional phase gate is employed.

\paragraph{Backward selection rule.}
Starting from the exact target \(|\Psi_{\mathrm{tar}}\rangle=|g\rangle\otimes|C_+(\alpha/2)\rangle\), at each \(n\) choose \(\theta_n\) so that the amplitude on \(|g,n\rangle\) is removed by \(S_n\):
\begin{equation}
\theta_n=\dfrac{\pi}{2}+\arctan \Big(\dfrac{a_{e,n-1}}{i a_{g,n}} \Big),
\qquad\Rightarrow\qquad a'_{g,n}=0.
\label{eq:theta-choice}
\end{equation}
Then apply a fixed single-qubit flip
\(Q_n = X_{\pi}=\exp(-i\tfrac{\pi}{2}\sigma_x)\)
within \(\{|g,n-1\rangle,|e,n-1\rangle\}\) to transfer the residual population to \(|g,n-1\rangle\) \cite{Law1996}.
Applying Eqs.~\eqref{eq:Sn-matrix}–\eqref{eq:theta-choice} for \(n=6,5,\dots,1\) gives a six-step backward sequence; the forward (experimental) order is the reverse:
\begin{equation}
Q_1\ \rightarrow\ S_1\ \rightarrow\ Q_2\ \rightarrow\ S_2\ \rightarrow\ \cdots\ \rightarrow\ Q_6\ \rightarrow\ S_6,
\qquad
t_n=\frac{\theta_n}{\sqrt{n}\,\xi}.
\end{equation}

\paragraph{ Truncated-subspace overlap.}
To quantify the error introduced by fixing the cutoff at \(N^\star=6\),
we evaluate the overlap between the ideal even cat and its projection
onto the truncated Fock subspace for the experimental parameter
\(\alpha=3.3\). The state used as input to the backward construction is
\begin{equation}
|\psi_6\rangle
=|g\rangle\otimes
\frac{\Pi_{\le 6}|C_+(\alpha/2)\rangle}{\sqrt{\langle C_+(\alpha/2)|\Pi_{\le 6}|C_+(\alpha/2)\rangle}},
\qquad
\Pi_{\le 6}=\sum_{n=0}^{6}|n\rangle\langle n|.
\end{equation}
For \(\alpha/2=1.65\), the nonvanishing (even) Fock amplitudes are
\begin{equation}
c_0\simeq0.36,\qquad c_2\simeq0.70,\qquad c_4\simeq0.55,\qquad c_6\simeq0.27.
\end{equation}
The probability weight retained in the truncated subspace is therefore
 \begin{equation}
P_{\le 6} = \langle C_+(\alpha/2)|\Pi_{\le 6}|C_+(\alpha/2)\rangle
=\sum_{n\in\{0,2,4,6\}}|c_n|^2
\simeq 0.989,
\end{equation}
so the neglected probability weight above \(n=6\) is about \( 1.07\%\). This quantifies the intrinsic truncation error.

 Table~\ref{tab:backward-sequence} compiles the six-step backward sequence (the time-reverse of the laboratory protocol for \(\alpha/2=1.65\)), listing the chosen angles \(\theta_n\) and the resulting system state after each operation; when read from top to bottom, it traces the population flow toward the vacuum, whereas the forward preparation is executed in the opposite order.

By construction of the angles in Table~\ref{tab:backward-sequence}, the backward-elimination rule ensures that applying the sequence in the order \(Q_1 S_1 \cdots Q_6 S_6\) to the target state recovers the vacuum:
\begin{equation}
Q_1 S_1 \cdots Q_6 S_6 \, |\psi_6\rangle \;=\; |\psi_0\rangle.
\label{eq:QSkill}
\end{equation}
Taking the adjoint and reversing the order gives the ideal preparation from vacuum to the target,
\begin{equation}
|\psi_6\rangle \;=\; S_6^\dagger Q_6^\dagger \cdots S_1^\dagger Q_1^\dagger \, |\psi_0\rangle.
\label{eq:ideal-dagger}
\end{equation}

For experimental convenience (to avoid excessively long or short swap durations on each \(n\)-manifold), we implement instead the laboratory “forward” composition
\begin{equation}
U_{\mathrm{fwd}} \;=\; S_6 Q_6 \cdots S_1 Q_1,
\qquad
|\psi_{\mathrm{fwd}}\rangle \;=\; U_{\mathrm{fwd}}\,|\psi_0\rangle.
\label{eq:Ufwd-def}
\end{equation}
We now show that this forward implementation also prepares \(|\psi_6\rangle\).

The laboratory swap $S_n=\exp(-iH_{\rm JC}t_n)$ is generated on the full
qubit--resonator Hilbert space by
\begin{equation}
H_{\rm JC}=\xi\big(a^\dagger\sigma_-+a\sigma_+\big),
\label{eq:SnQn}
\end{equation}
and acts simultaneously on all excitation manifolds. Define the global
qubit operator
\begin{equation}
Z_q=\big(|g\rangle\!\langle g|-|e\rangle\!\langle e|\big)\otimes\mathbb I_{\rm cav}.
\label{eq:Zn}
\end{equation}
It obeys $Z_q^2=\mathbb I$, $Z_qH_{\rm JC}Z_q=-H_{\rm JC}$, and
$Z_q\sigma_xZ_q=-\sigma_x$. Since $Q_n=X_\pi=-i\sigma_x$, this gives
\begin{equation}
Z_q S_n Z_q=S_n^\dagger,
\qquad
Z_q Q_n Z_q=Q_n^\dagger.
\label{eq:conj}
\end{equation}
Using Eq. \eqref{eq:conj} repeatedly, we obtain
\begin{equation}
S_6^\dagger Q_6^\dagger \cdots S_1^\dagger Q_1^\dagger
\;=\;
Z_q\,\big(S_6 Q_6 \cdots S_1 Q_1\big)\,Z_q
\;=\;
Z_q\,U_{\mathrm{fwd}}\,Z_q.
\label{eq:Ubwd-Ufwd}
\end{equation}
Since both endpoints lie in the ground-qubit sector,
\(Z_q|\psi_0\rangle=|\psi_0\rangle\) and \(Z_q|\psi_6\rangle=|\psi_6\rangle\).
Combining Eqs.~\eqref{eq:ideal-dagger} and \eqref{eq:Ubwd-Ufwd}, and then
multiplying by $Z_q$, gives
\begin{equation}
|\psi_6\rangle=Z_qU_{\mathrm{fwd}}|\psi_0\rangle,
\qquad\Longrightarrow\qquad
U_{\mathrm{fwd}}|\psi_0\rangle=Z_q|\psi_6\rangle=|\psi_6\rangle.
\label{eq:final-rel-short}
\end{equation}
Thus the forward composition prepares the target exactly. Any sign differences induced by \(Z_q\) affect only intermediate \(|e\rangle\otimes|n\!-\!1\rangle\) components and are absent at the endpoints.

\begin{figure}[t]
\centering
    \includegraphics[width=0.95\linewidth]{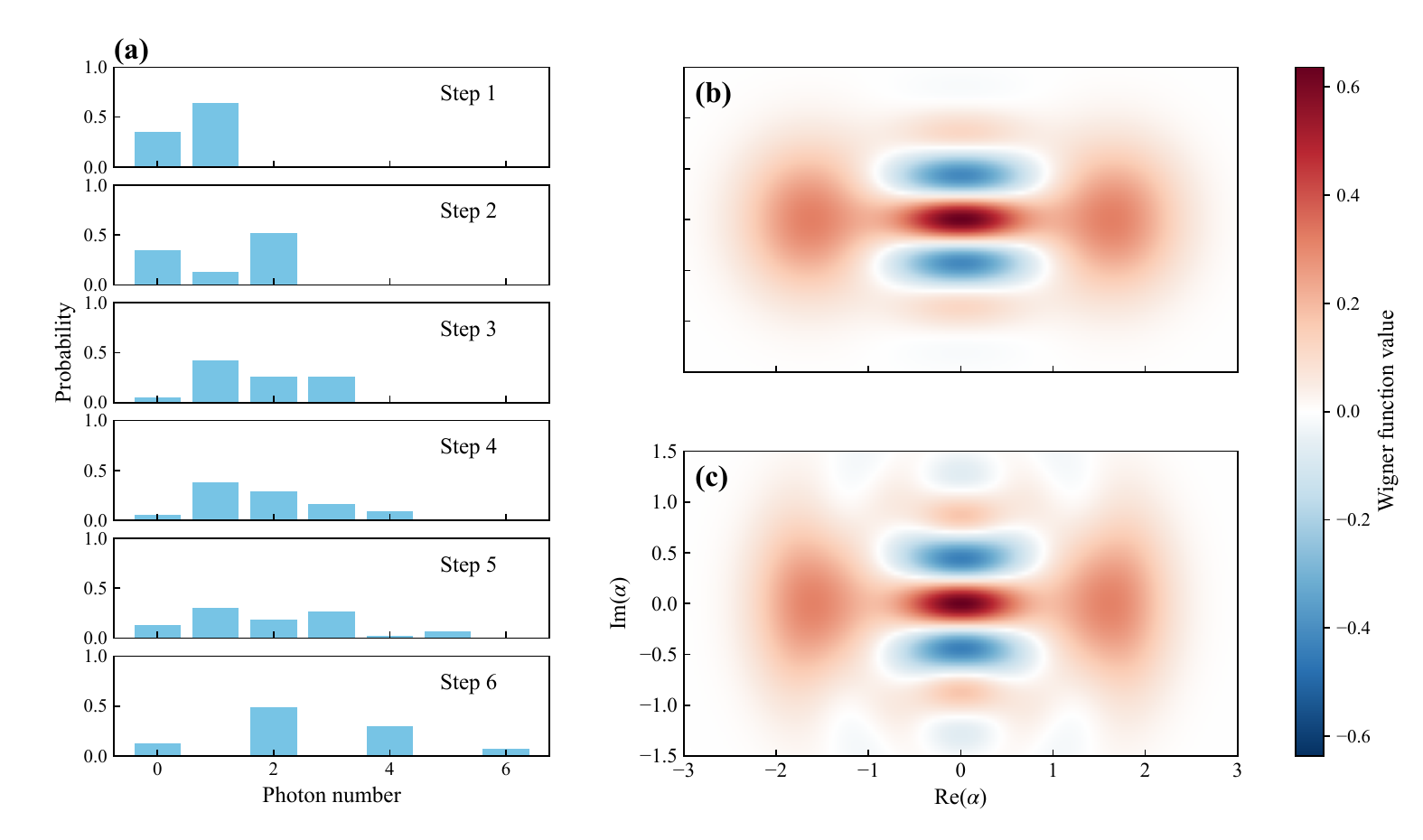}
  \caption{ \textbf{Forward six-step protocol.}
(a) Photon-number distributions (\(n=0\!-\!6\)) for the intermediate resonator
states obtained by the six-step protocol in Table~\ref{tab:backward-sequence}; panels from top to bottom correspond
to the forward steps 1–6, showing how population is progressively transferred from lower
to higher \(n\)-excitation manifolds. Reversing the tabulated operations gives the backward
sequence used for analysis. (b) Wigner function of the ideal even cat
\(|C_{+}(\alpha/2)\rangle\) with \(\alpha/2=1.65\) (\(\alpha=3.3\)). (c) Wigner function computed from the theoretical state \(|\psi_6\rangle\) produced by the forward sequence.
The color bar indicates the Wigner-function value; axes are labeled by \(\mathrm{Re}\,\alpha\) and \(\mathrm{Im}\,\alpha\).
}
\label{fig:steps-figure}
\end{figure}

Figure~\ref{fig:steps-figure}(a) visualizes the evolution in the forward direction: panels from top to bottom correspond to steps 1–6 of the laboratory sequence and show how population is redistributed among the \(n\)-excitation manifolds as the protocol proceeds.
For comparison, Fig.~\ref{fig:steps-figure}(b) shows the Wigner function of the ideal even cat \(|C_{+}(\alpha/2)\rangle\) with \(\alpha/2=1.65\), while Fig.~\ref{fig:steps-figure}(c) presents the Wigner function computed from the theoretical state \(|\psi_6\rangle\) produced by the forward sequence, whose close agreement confirms the effectiveness of the protocol.

\begin{table}[h!]
\centering
\label{tab:backward-sequence}
\begin{tabular}{l@{\hskip 2cm}c@{\hskip 2cm}l}
\hline
Sequence of states, & Operational & \multirow{2}{*}{System state } \\
 operations & parameter & \\
\hline
\(|\psi_6\rangle\) & & \(|g\rangle \otimes\big(0.36|0\rangle+0.70|2\rangle+0.55|4\rangle+0.27|6\rangle\big)\) \\
\(S_6\) & \(\theta_6=1.57\) & \\
\(Q_6\) & \(X_{\pi}\) & \\
\multirow{2}{*}{\(|\psi_5\rangle\)} & &
\multirow{2}{*}{ \( \begin{aligned}
|g\rangle\otimes(-0.55|1\rangle-0.52|3\rangle-0.27|5\rangle) \\
-i|e\rangle\otimes(0.36|0\rangle+0.43|2\rangle+0.16|4\rangle)
\end{aligned} \)} \\ \\
\hline
\(S_5\) & \(\theta_5=2.09\) & \\
\(Q_5\) & \(X_{\pi}\) & \\
\multirow{2}{*}{\(|\psi_4\rangle\)} & & \multirow{2}{*}{ \( \begin{aligned}
&|g\rangle\otimes(0.23|0\rangle+0.54|2\rangle+0.31|4\rangle)\\
&+i|e\rangle\otimes(0.62|1\rangle+0.40|3\rangle)
\end{aligned} \)} \\ \\
\hline
\(S_4\) & \(\theta_4=2.48\) & \\
\(Q_4\) & \(X_{\pi}\) & \\
\multirow{2}{*}{\(|\psi_3\rangle\)} & & \multirow{2}{*}{ \(
\begin{aligned}
|g\rangle\otimes(-0.65|1\rangle-0.51|3\rangle) \\
-i|e\rangle\otimes(0.23|0\rangle+0.51|2\rangle)
\end{aligned} \)} \\ \\
\hline
\(S_3\) & \(\theta_3=2.35\) & \\
\(Q_3\) & \(X_{\pi}\) & \\
\(|\psi_2\rangle\) & & \( |g\rangle\otimes(0.59|0\rangle+0.72|2\rangle) +i|e\rangle\otimes(0.36|1\rangle)
\) \\
\hline
\(S_2\) & \(\theta_2=2.03\) & \\
\(Q_2\) & \(X_{\pi}\) & \\
\(|\psi_1\rangle\) & & \( |g\rangle\otimes(-0.80|1\rangle)
-i|e\rangle\otimes(0.59|0\rangle )
\) \\
\hline
\(S_1\) & \(\theta_1=2.20\) & \\
\(Q_1\) & \(X_{\pi}\) & \\
\(|\psi_0\rangle\) & & \(|g\rangle \otimes |0\rangle\) \\
\hline
\end{tabular}
\caption{\textbf{Six-step backward sequence for \(\bm{\alpha/2=1.65}\).}
Entries list the swap angle \(\theta_n\) (for \(S_n\)) and the resulting system state after each operation. Rows are given in backward order; the laboratory implementation applies them from bottom to top (forward order).}
\end{table}

\subsection{From a phase cat to an amplitude cat via cavity displacement}

Building on the six-step protocol and its backward verification in Fig.~\ref{fig:steps-figure}, we next convert the prepared phase cat \(|C_{+}(\alpha/2)\rangle\) into a standard amplitude cat by a calibrated cavity displacement.

\begin{figure*}[t]
\centering
    \includegraphics[width=1\linewidth]{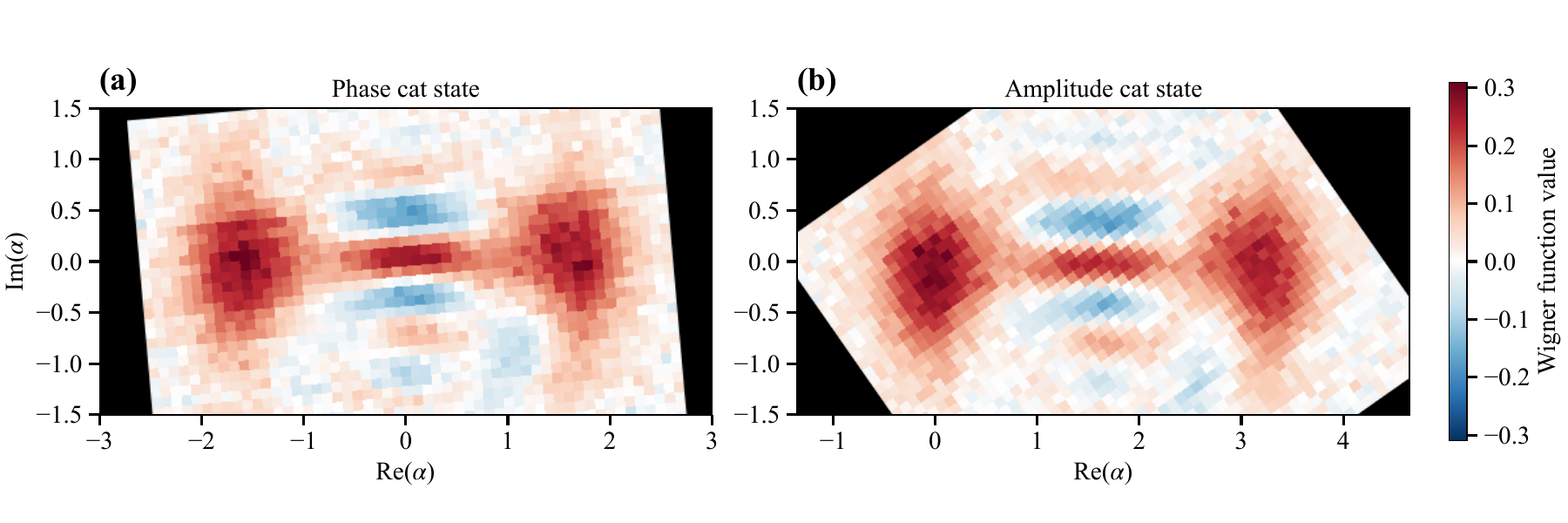}
  \caption{
\textbf{Displacing a phase cat to an amplitude cat.}
(a) Wigner tomography of the phase cat prepared by the photon-by-photon protocol,
\(\mathcal N_{+} (|\alpha/2\rangle+|-\alpha/2\rangle )\).
(b) After a resonant displacement \(D(\alpha/2)\), the coherent components move to \(0\) and \(\alpha\), yielding the amplitude cat \(\mathcal N_{+}(|0\rangle+|\alpha\rangle)\).
Small phase-space rotations arise because the AC-Stark shifts in the experimental rotating frame and in the tomography rotating frame are not identical, producing a net angle \(\vartheta\) described by \(R(\vartheta)=e^{-i\vartheta a^\dagger a}\).
All Wigner plots are de-rotated for clarity.
}
\label{fig:cat-displacement}
\end{figure*}

The state in Fig.~\ref{fig:cat-displacement}(a) is the even phase cat
\begin{equation}
|C_{+}(\alpha/2)\rangle
= \mathcal N_{+}\,\big(|\alpha/2\rangle+|-\alpha/2\rangle\big),
\end{equation}
where \(\mathcal N_{+}\) is the normalization factor.
To obtain an amplitude cat of the form \(|0\rangle+|\alpha\rangle\), we apply the displacement operator
\begin{equation}
D(\beta)=\exp\!\big(\beta a^\dagger-\beta^* a\big),\qquad
D(\beta)\,|\gamma\rangle = e^{i\,\mathrm{Im}(\beta\gamma^*)}\,|\gamma+\beta\rangle.
\end{equation}
Choosing \(\beta=\alpha/2\) gives,
\begin{equation}
D(\alpha/2)\,|C_{+}(\alpha/2)\rangle
= \mathcal N_{+}\,\big(|\alpha\rangle+|0\rangle\big).
\end{equation}

Experimentally, \(D(\beta)\) is realized by a short resonant microwave pulse applied on the cavity with drive Hamiltonian
\begin{equation}
H_d(t)=\hbar\big(\epsilon a^\dagger e^{-i\omega_s t}+\epsilon^* a e^{i\omega_s t}\big),
\end{equation}
where \(\omega_s/(2\pi)\) denotes the cavity frequency.
In the cavity frame rotating at \(\omega_s\), this reduces to
\(\tilde H_d(t)=\hbar(\epsilon a^\dagger+\epsilon^* a)\) and produces a net displacement
\begin{equation}
U(t)=\mathcal T\exp\!\Big[-\frac{i}{\hbar}\!\int_0^t \tilde H_d(\tau)\,d\tau\Big]
= D(\beta),\qquad
\beta=-i\!\int_0^t \epsilon(\tau)\,d\tau.
\end{equation}
By setting \(|\beta|=|\alpha|/2\) and aligning \(\arg\beta\) along the real axis, the phase cat \(|C_{+}(\alpha/2)\rangle\) is translated into the amplitude cat \(\mathcal N_{+}(|0\rangle+|\alpha\rangle)\).

In Fig.~\ref{fig:cat-displacement} we display the Wigner functions of the prepared phase and amplitude cats.
For clarity, these plots are shown after a small de-rotation applied to the measured states,
since in practice the experimentally measured Wigner functions can appear rotated in phase space.
This rotation originates from a mismatch between the AC-Stark shifts accumulated in the experimental rotating frame
and those in the tomography rotating frame, which can be modeled as an additional effective term \(\hbar\delta\chi\, a^\dagger a\).
The accumulated difference \(\vartheta=\int\delta\chi\,dt\) implements a phase-space rotation
\(R(\vartheta)=e^{-i\vartheta a^\dagger a}\).
We de-rotate the measured states before plotting, so that the Wigner functions displayed in Fig.~\ref{fig:cat-displacement} correspond to the intended cat states rather than their rotated versions.

 \paragraph{Cat-state fidelity from Rabi reconstruction.}
\label{sec:cat-fidelity-coherence}

We reconstruct the phase- and amplitude-cat density matrices from the calibrated Rabi records using a dissipative three-level Jaynes--Cummings readout model. The reconstructions are constrained by positive semidefiniteness and unit trace, with the purities fixed to the independently estimated values ${\rm Tr}(\rho_{\rm phase}^{2})=0.823$ and ${\rm Tr}(\rho_{\rm amplitude}^{2})=0.789$. These purities are estimated by simulating the known preparation sequences using the measured qubit and resonator coherence parameters. For the amplitude cat, we undo the calibrated final displacement $D(\alpha/2)$ and perform the reconstruction in the resulting low-Fock frame.

We compare the reconstructed states with ideal cats using the Uhlmann fidelity~\cite{NielsenChuang2000}
\begin{equation}
F_{\rm U}(\rho,\sigma)
={\rm Tr}\sqrt{\sqrt{\rho}\,\sigma\sqrt{\rho}},
\qquad
F_{\rm U}\!\left(\rho,|\mathcal C\rangle\langle\mathcal C|\right)
=\sqrt{\langle\mathcal C|\rho|\mathcal C\rangle}.
\label{eq:uhlmann-cat-fidelity}
\end{equation}
We obtain fidelities of $0.929$ and $0.913$ for the phase and amplitude cats, respectively.

 \section{Calibration of Qubit Z-Line Signals}
Before turning to the specific calibration routines, we note that the purpose of characterizing qubit Z-line signals is to ensure that subsequent Floquet-type modulations are both accurate and reproducible.
Because sideband coupling via Floquet modulation relies on precise periodic driving through the Z lines, any static miscalibration of Z-line amplitudes, waveform distortions, or inter-qubit crosstalk would directly degrade the modulation accuracy.
Therefore, in this section we establish a set of calibrations for the Z-line signals of all qubits, including amplitude-to-frequency mapping, verification of square-wave integrity, and compensation of Z-line crosstalk.
These steps guarantee that the Z-line drive serves as a reliable and calibrated control channel for the Floquet protocols introduced later.

\subsection{Mapping Z-line Square Pulse Amplitude to Qubit Frequency}

\begin{figure}[htbp]
    \centering
    \includegraphics[width=0.9\linewidth]{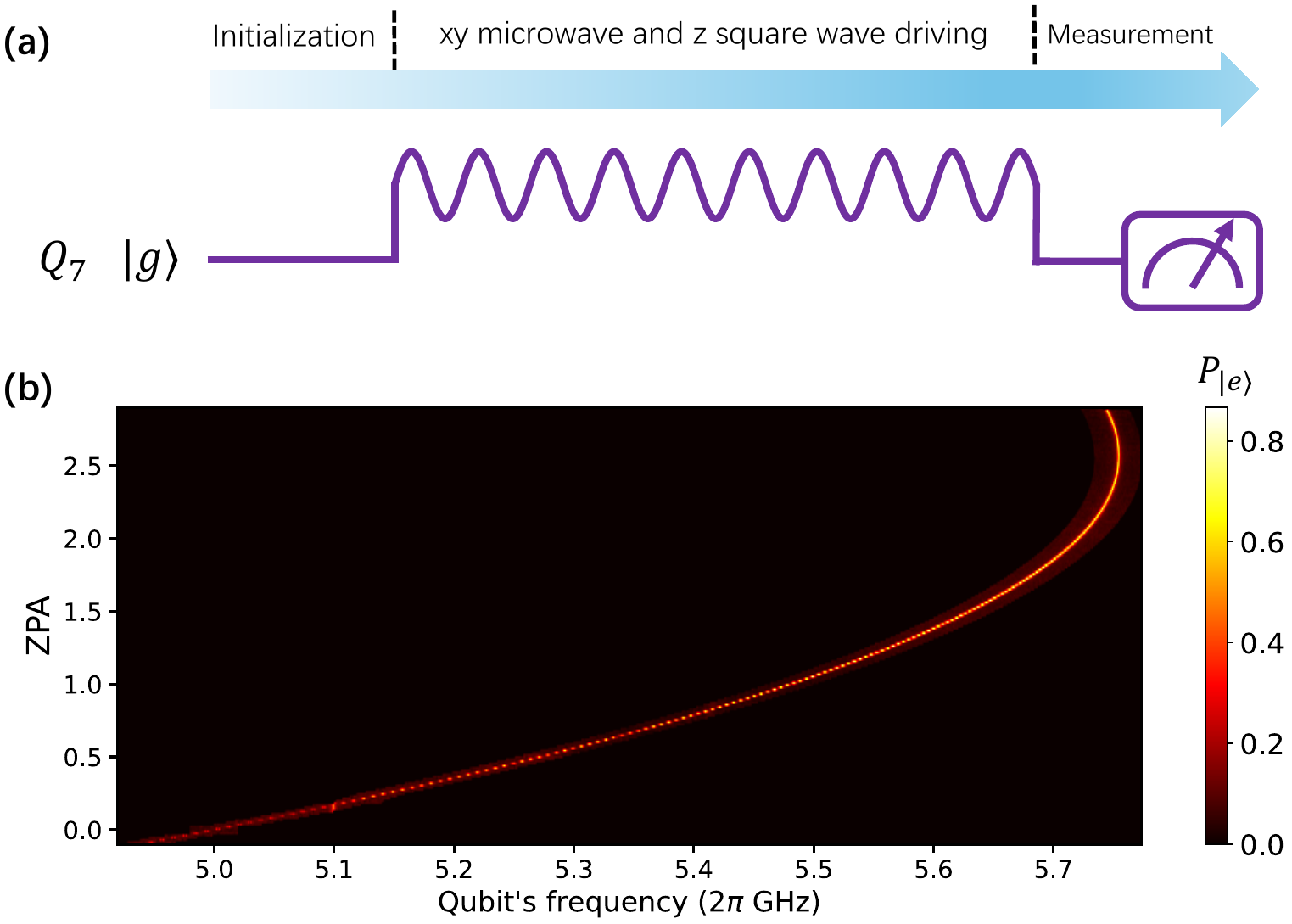}
    \caption{\textbf{Mapping between ZPA and the qubit frequency.}
    (a) Pulse sequence: initialization in $|g\rangle$, concurrent XY-line microwave tone and Z-line square wave, followed by measurement. The ZPA sets the qubit frequency; the XY tone probes resonance.
    (b) Excited-state population versus XY-drive frequency and ZPA; the bright resonance ridge yields $f_q(\mathrm{ZPA})$. The same protocol is applied to all qubits to obtain per-qubit calibrations.}
    \label{fig:R7-ZPA}
\end{figure}

In this calibration, we determine the function $f_q(\mathrm{ZPA})$ that maps the Z-line square-pulse amplitude (ZPA) to the qubit frequency. The Z line is driven by a square wave whose flat-top amplitude sets the qubit frequency, while a microwave tone on the XY line is used as a spectroscopy probe. The pulse sequence used for this procedure is illustrated in Fig.~\ref{fig:R7-ZPA}(a): initialize the qubit in $|g\rangle$, then apply a concurrent XY microwave tone and a Z-line square wave, followed by a projective measurement. We present qubit $Q_7$ as a representative example; the same protocol is applied to all other qubits to obtain their individual $f_q^{(i)}(\mathrm{ZPA})$.

For a fixed ZPA, we sweep the XY-drive frequency concurrently with the Z-line square wave and record the excited-state population. The resonance peak gives the qubit frequency at that ZPA. Repeating this over a range of ZPAs yields the two-dimensional map shown in Fig.~\ref{fig:R7-ZPA}(b) (excited-state population $P_{|e\rangle}$ versus XY-drive frequency and ZPA). The bright ridge traces $f_q(\mathrm{ZPA})$, which we extract by fitting the ridge with a polynomial fit. The resulting calibration is then used for precise frequency placement via Z-line control in subsequent experiments. The same mapping-and-fit workflow is performed for each remaining qubit to build a per-qubit lookup table and to update it as needed prior to data acquisition.

\subsection{Verification of Z-Pulse Square-Wave Integrity}

We verify the time-domain integrity of the Z square pulse by fixing the XY-line microwave at a constant amplitude [on-resonance Rabi rate $\Omega^{Z}/(2\pi) \approx 3~\mathrm{MHz}$] and a fixed drive frequency $\omega_d^{Z}/(2\pi)=5.6~\mathrm{GHz}$.
For each ZPA, the Z-line square wave shifts the qubit's angular frequency $\omega_q^Z$, thereby yielding a detuning in the drive frame
\begin{equation}
\delta^{Z}(t) = \omega_q^Z(t) - \omega_d^{Z},
\label{eq:detuningZ}
\end{equation}
where the superscript $Z$ denotes quantities defined for this Z-pulse verification (the ZPA dependence is implicit below).
The pulse sequence is analogous to the mapping sequence of Fig.~\ref{fig:R7-ZPA}(a): initialize the qubit in $|g\rangle$, apply a concurrent XY-line microwave tone (fixed amplitude and frequency) and a Z-line square wave, and then measure the qubit state as a function of the evolution time $t$.
Panel (b) records the resulting Rabi-oscillation traces versus time for different ZPA values, and panel (c) shows their Fourier spectra taken along the time axis.

\begin{figure}[htbp]
    \centering
    \includegraphics[width=\linewidth]{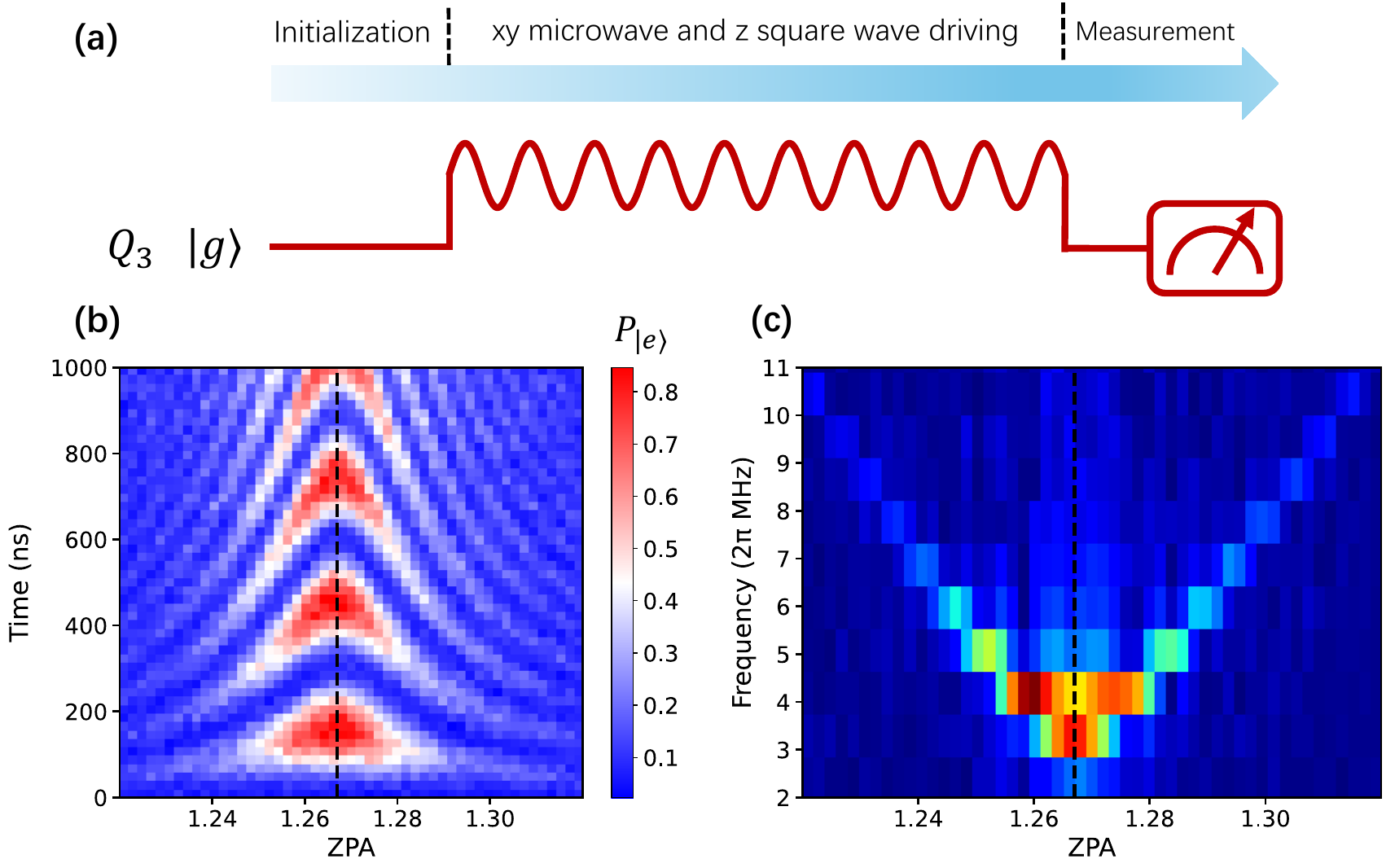}
    \caption{\textbf{Verification of Z-pulse square-wave integrity.}
    (a) Pulse sequence for the verification, analogous to the ZPA$\rightarrow$frequency mapping in Fig.~\ref{fig:R7-ZPA}(a): initialization in $|g\rangle$, concurrent XY-line microwave tone (fixed amplitude and frequency) and Z-line square wave, followed by measurement.
    (b) Rabi-oscillation traces versus time for different ZPA values; the dashed line marks the ZPA at which $\delta^{Z}=0$ for selected times.
    (c) Fourier spectrum (computed along $t$) versus ZPA. The V-shaped ridge follows $\Omega^{Z}_R/(2\pi)$ [Eq.~\eqref{eq:OmegaRZ}], minimizing at $\Omega^{Z}/2\pi$ when $\delta^{Z}=0$. The time-dependent locus of minimal frequency reveals the temporal shape imparted by the Z pulse and is used to predistort the waveform to an ideal square.}
    \label{fig:ZSquareVerify}
\end{figure}

In the rotating frame, the driven two-level Hamiltonian is
\begin{equation}
H(t)=\frac{1}{2}\,\delta^{Z}(t)\,\sigma_z+\frac{1}{2}\,\Omega^{Z}\,\sigma_x,
\label{eq:HdriveZ}
\end{equation}
with the generalized Rabi frequency
\begin{equation}
\Omega^{Z}_R(t)=\sqrt{(\Omega^{Z})^{2}+ \big[\delta^{Z}(t)\big]^2 }.
\label{eq:OmegaRZ}
\end{equation}
Starting from $|g\rangle$, the excited-state probability reads
\begin{equation}
P_{|e\rangle}(t) \approx \frac{(\Omega^{Z})^{2}}{\big[\Omega^{Z}_R(t)\big]^2}\,
\sin^{2}\!\Big(\frac{1}{2}\Omega^{Z}_R(t)\,t\Big).
\label{eq:PeZ}
\end{equation}
Hence the oscillation frequency along the time axis equals $\Omega^{Z}_R(t)/(2\pi)$: it is minimal at zero detuning ($\delta^{Z}=0$) and increases monotonically with $|\delta^{Z}|$.
For each ZPA we Fourier-transform the time traces to obtain Fig.~\ref{fig:ZSquareVerify}(c); the V-shaped spectral ridge follows $\Omega^{Z}_R/(2\pi)$ with its minimum at $\Omega^{Z}/2\pi\approx 3~\mathrm{MHz}$ when $\delta^{Z}=0$.
By locating, for each time slice, the ZPA where this spectral peak is minimal, we obtain the locus $\delta^{Z}(t)=0$, which vertically outlines how the Z pulse sets the qubit frequency throughout the pulse.
An ideal square wave yields a flat (vertical) trace; any tilt or curvature indicates distortion (e.g., capacitive filtering).
We use this diagnostic to predistort the Z waveform (e.g., exponential or Gaussian edge compensation) until the zero-detuning trace is flat.
The same verification-and-correction procedure is applied to all qubits.

\subsection{Calibration of Qubit Z-Line Crosstalk}

\begin{figure}[htbp]
    \centering
    \includegraphics[width=0.95\linewidth]{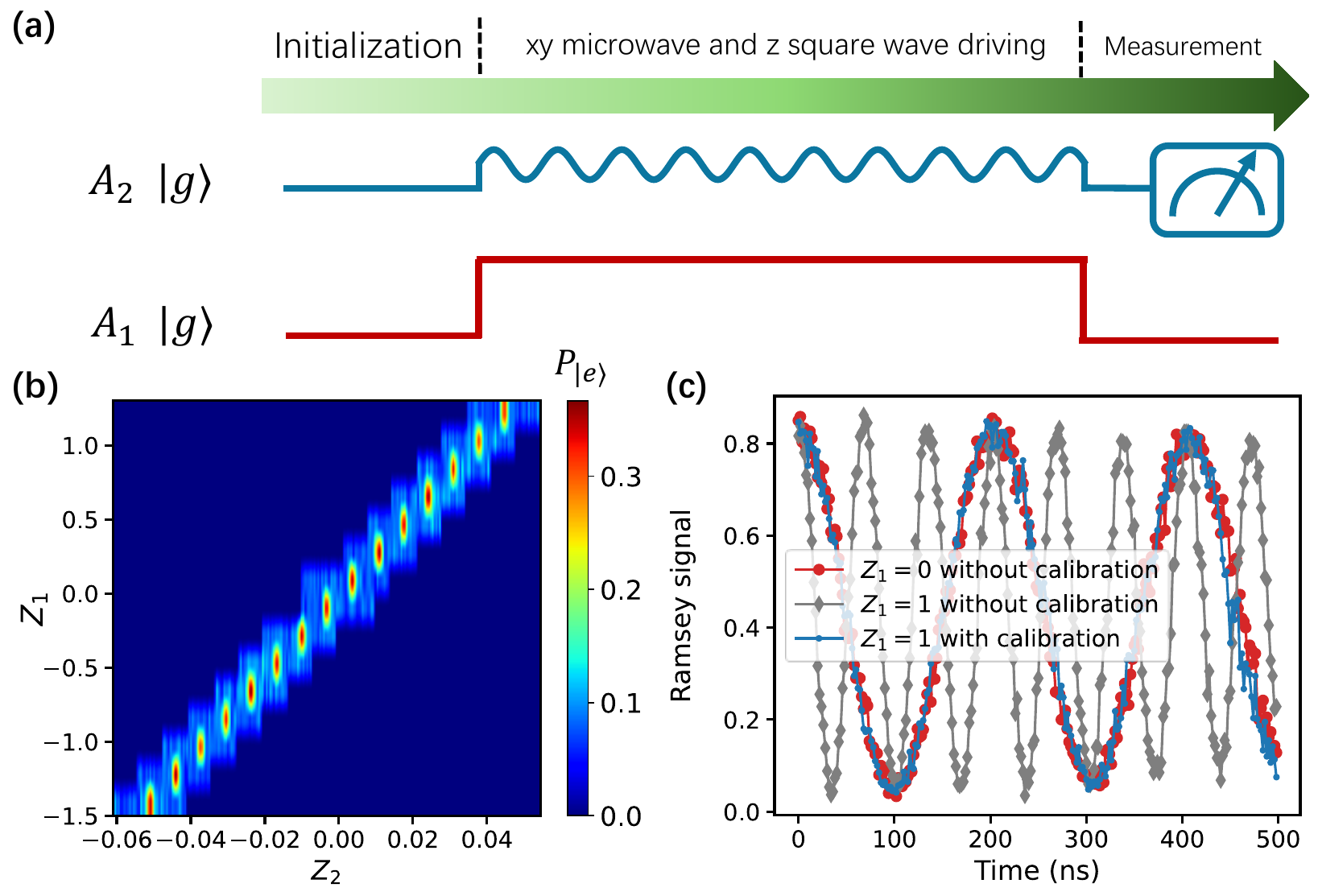}
    \caption{\textbf{Calibration of the Z crosstalk.}
    (a) Pulse sequence. Apply a Z-line square wave of amplitude \(Z_{1}\) to \(A_{1}\) and, simultaneously, a small Z square wave of amplitude \(Z_{2}\) to \(A_{2}\); keep an XY-line microwave on \(A_{2}\) at its idle frequency; then measure \(A_{2}\).
    (b) Excited-state population of \(A_{2}\) versus \(Z_{1}\) and \(Z_{2}\). The bright ridge gives \(Z_{2}=\alpha^{Z}_{21} Z_{1}\).
    (c) Ramsey verification on \(A_{2}\) with preset baseline \(f_{R}^{\mathrm{base}} =5~\mathrm{MHz}\): comparison among \(Z_{1}=0\) (baseline), \(Z_{1}=1\) without compensation (frequency shifted), and \(Z_{1}=1\) with compensation \(Z_{2}=\alpha^{Z}_{21} Z_{1}\) (frequency restored).}
    \label{fig:Zcross}
\end{figure}

Because the Z lines are routed in close proximity, driving one qubit can induce a smaller square-wave component on another. When a Z square wave with amplitude \(Z_{1}\) is applied to \(A_{1}\), a parasitic input appears on the Z line of \(A_{2}\). To cancel it, we apply a compensation square wave with amplitude \(Z_{2}\) on \(A_{2}\), while simultaneously driving \(A_{2}\) on its XY line at the idle frequency. By scanning \(Z_{1}\) and \(Z_{2}\) and measuring the excited-state population of \(A_{2}\), we obtain a two-dimensional map [Fig.~\ref{fig:Zcross}(b)] whose bright ridge yields the linear relation
\begin{equation}
Z_{2} \;=\; \alpha^{Z}_{21}\, Z_{1},
\label{eq:zxt_linear}
\end{equation}
where \( \alpha^{Z}_{21} \) quantifies the Z crosstalk from \(A_{1}\!\to\!A_{2}\). In subsequent experiments, for any commanded \(Z_{1}\) we set \(Z_{2}\) according to Eq.~\eqref{eq:zxt_linear} so that the frequency of \(A_{2}\) remains fixed.

\paragraph{Ramsey verification.}
We further verify the calibration with Ramsey interferometry on \(A_{2}\). Two \(\pi/2\) pulses on the XY line of \(A_{2}\), separated by a variable delay, are applied while the Z square waves are present. The XY carrier is detuned from the idle frequency by a preset value
\begin{equation}
f_{R}^{\mathrm{base}}=5~\mathrm{MHz},
\label{eq:ramsey_set}
\end{equation}
so that in the absence of crosstalk the Ramsey fringes oscillate at \(f_{R}^{\mathrm{base}}\).
If the Z pulse on \(A_{1}\) shifts the frequency of \(A_{2}\) by \(\delta^{Z}_{21}\), the observed Ramsey frequency becomes
\begin{equation}
f_{R} = f_{R}^{\mathrm{base}} + \delta^{Z}_{21}.
\label{eq:ramsey_freq}
\end{equation}
We compare three cases:
(i) \(Z_{1}=0\) (baseline), yielding \(f_{R}=f_{R}^{\mathrm{base}}\);
(ii) \(Z_{1}=1\) without compensation, giving a shifted frequency \(f_{R}^{\mathrm{ct}}\neq f_{R}^{\mathrm{base}}\);
(iii) \(Z_{1}=1\) with compensation \(Z_{2}=\alpha^{Z}_{21} Z_{1}\), which restores the Ramsey frequency to \(f_{R}^{\mathrm{base}}\).
The recovery of the 5-MHz fringes demonstrates that the Z crosstalk has been effectively canceled. This protocol can be repeated for any pair of qubits to obtain the full set of crosstalk coefficients \(\alpha^{Z}_{ij}\).

Once the pairwise coefficients $\alpha^{Z}_{ij}$ are measured, we arrange the Z amplitudes as column vectors
\begin{equation}
\mathbf{Z}^{\mathrm{cmd}} =
\begin{bmatrix}
Z^{\mathrm{cmd}}_{1} \\[2pt]
Z^{\mathrm{cmd}}_{2} \\[2pt]
\vdots \\[2pt]
Z^{\mathrm{cmd}}_{N}
\end{bmatrix},
\qquad
\mathbf{Z}^{\mathrm{eff}} =
\begin{bmatrix}
Z^{\mathrm{eff}}_{1} \\[2pt]
Z^{\mathrm{eff}}_{2} \\[2pt]
\vdots \\[2pt]
Z^{\mathrm{eff}}_{N}
\end{bmatrix},
\end{equation}
where $Z^{\mathrm{cmd}}_{j}$ is the commanded square-wave amplitude put on the physical Z line of qubit $j$, and $Z^{\mathrm{eff}}_{i}$ is the amplitude effectively experienced by qubit $i$ after crosstalk.

The two are connected by the correction matrix $M_{\mathrm{cor}}$,
\begin{equation}
\mathbf{Z}^{\mathrm{eff}} = M_{\mathrm{cor}}\,\mathbf{Z}^{\mathrm{cmd}},
\end{equation}
whose explicit form (for $N$ qubits) is
\begin{equation}
M_{\mathrm{cor}} =
\begin{bmatrix}
1 & -\alpha^{Z}_{12} & -\alpha^{Z}_{13} & \cdots & -\alpha^{Z}_{1N} \\
-\alpha^{Z}_{21} & 1 & -\alpha^{Z}_{23} & \cdots & -\alpha^{Z}_{2N} \\
-\alpha^{Z}_{31} & -\alpha^{Z}_{32} & 1 & \cdots & -\alpha^{Z}_{3N} \\
\vdots & \vdots & \vdots & \ddots & \vdots \\
-\alpha^{Z}_{N1} & -\alpha^{Z}_{N2} & -\alpha^{Z}_{N3} & \cdots & 1
\end{bmatrix},
\end{equation}
with diagonal entries 1 (local action) and off-diagonal entries \(-\alpha^{Z}_{ij}\) [parasitic contribution from $j$ to $i$, extracted from the compensation ridge in Fig.~\ref{fig:Zcross}(b)].

Given a desired effective configuration
\(\mathbf{Z}^{\mathrm{eff}}_{\!*}\),
the commanded amplitudes are obtained by solving the linear system
\begin{equation}
M_{\mathrm{cor}}\,\mathbf{Z}^{\mathrm{cmd}}=\mathbf{Z}^{\mathrm{eff}}_{\!*}
\quad\Rightarrow\quad
\mathbf{Z}^{\mathrm{cmd}}=M_{\mathrm{cor}}^{-1}\,\mathbf{Z}^{\mathrm{eff}}_{\!*}.
\end{equation}
Operationally, once $M_{\mathrm{cor}}$ is calibrated, each programmed Z waveform is pre-multiplied by $M_{\mathrm{cor}}^{-1}$ before being sent to the hardware, so that the qubits receive the intended effective Z amplitudes during experiments.

\section{Sideband Coupling via Floquet Modulation}

We consider an Xmon qubit \(Q_j\) dispersively coupled to a cavity mode.
The cavity annihilation (creation) operator is \(a\) (\(a^{\dagger}\)) with angular frequency \(\omega_{s}\).
The symbol \(q\) (\(q^{\dagger}\)) denotes the annihilation (creation) operator of the anharmonic oscillator mode of qubit \(Q_j\), whose mean angular frequency is \(\omega_j^m\).
    All sideband interactions are operated at nominal resonance.

To realize the first-order sideband, we set the qubit angular frequency to
\begin{equation}
\omega_j^m=\omega_{s}-\nu_j.
\label{eq:sb-ws-def}
\end{equation}
A sinusoidal frequency modulation is applied to the qubit frequency. The total Hamiltonian is ($\hbar=1$)
\begin{align}
H &= H_0 + H_I, \label{eq:sb-H}\\
H_0 &= \omega_{s}\,a^{\dagger}a
      + \big[\omega_j^m + \varepsilon_j\cos(\nu_j t)\big]\,q^{\dagger}q
      - \frac{K_j}{2}\,q^{\dagger 2}q^{2}, \label{eq:sb-H0}\\
H_I &= \xi_j\,a^{\dagger}q + \text{h.c.}\,.
\label{eq:sb-HI}
\end{align}
Here \(\varepsilon_j\) and \(\nu_j\) are the amplitude and angular frequency of the qubit-frequency modulation, \(\xi_j\) is the coupling strength between \(Q_j\) and the cavity, and \(K_j\) is the anharmonicity of \(Q_j\).

We keep the lowest three levels \(\{|g\rangle,|e\rangle,|f\rangle\}\) of \(Q_j\) and move to the rotating frame of \(H_0\) with
\begin{equation}
	U_0(t)=\exp\Bigg\{ i\omega_s a^\dagger a\, t
	+i\left[\omega_j^m t+\frac{\varepsilon_j}{\nu_j}\sin(\nu_j t)\right]|e\rangle\langle e|
	+i\left[(2\omega_j^m-K_j)t+2\frac{\varepsilon_j}{\nu_j}\sin(\nu_j t)\right]|f\rangle\langle f|
	\Bigg\}.
    \label{eq:sb-U0}
\end{equation}
Defining the modulation index \(\mu_j= \varepsilon_j/\nu_j\), the interaction in this frame is
\begin{equation}
H'_I=
\Big\{
e^{-i\mu_j\sin(\nu_j t)}\,\xi_j\,a^{\dagger}
\big[\,e^{i\nu_j t}\,|g\rangle\!\langle e|
+\sqrt{2}\,e^{i(\nu_j+K_j)t}\,|e\rangle\!\langle f|\,\big]
+\text{h.c.}
\Big\}.
\label{eq:sb-Hprime}
\end{equation}

Using the Jacobi--Anger identity,
\begin{equation}
e^{-i\mu_j\sin(\nu_j t)}=\sum_{n=-\infty}^{\infty}J_{n}(\mu_j)\,e^{-in\nu_j t},
\label{eq:sb-JA}
\end{equation}
Equation~\eqref{eq:sb-Hprime} becomes
\begin{equation}
H'_I=
\sum_{n=-\infty}^{\infty}J_{n}(\mu_j)\,\xi_j\,a^{\dagger}
\Big[e^{-i(n-1)\nu_j t}|g\rangle\!\langle e|
+\sqrt{2}\,e^{-i(n-1)\nu_j t+iK_j t}|e\rangle\!\langle f|\Big]
+\text{h.c.}
\label{eq:sb-Hprime-J}
\end{equation}
The factor \((n-1)\nu_j\) shows that the \(n=1\) Floquet harmonic is stationary, which realizes the first-order sideband.

Assuming \(\{|J_{0}(\mu_j)\xi_j|\}\ll \nu_j\) and working close to the \(n=1\) sideband resonance, we keep the slowly varying \(n=1\) term in the \(|g\rangle\!\langle e|\) channel and treat the other harmonics dispersively to second order. The effective Hamiltonian is
\begin{equation}
H_{\mathrm{eff}}=
\Big[\,J_{1}(\mu_j)\,\xi_j\,a^{\dagger}|g\rangle\!\langle e|+\text{h.c.}\,\Big]
+ S_{1}\big(|g\rangle\!\langle g|-|e\rangle\!\langle e|\big)\,a^{\dagger}a
- S_{1}\,|e\rangle\!\langle e|
+ S_{2}\,|e\rangle\!\langle e|\,a^{\dagger}a.
\label{eq:sb-Heff-correct}
\end{equation}
The Stark shifts induced by off-resonant harmonics \(n\neq1\) are
\begin{equation}
S_1=\sum_{n\neq 1}\frac{\big[J_n(\mu_j)\,\xi_j\big]^2}{(1-n)\,\nu_j},
\qquad
S_2=\sum_{n=-\infty}^{\infty} \frac{2\big[J_n(\mu_j)\,\xi_j\big]^2}{(1-n)\,\nu_j+K_j}.
\label{eq:sb-S12-general-correct}
\end{equation}
Therefore the first-order sideband coupling strength is
\begin{equation}
\frac{\lambda_j}{2}=J_{1}(\mu_j)\,\xi_j.
\label{eq:sb-gSB}
\end{equation}

\subsection{Experimental demonstration of sideband coupling}

\begin{figure}[t]
    \centering
    \includegraphics[width=0.95\linewidth]{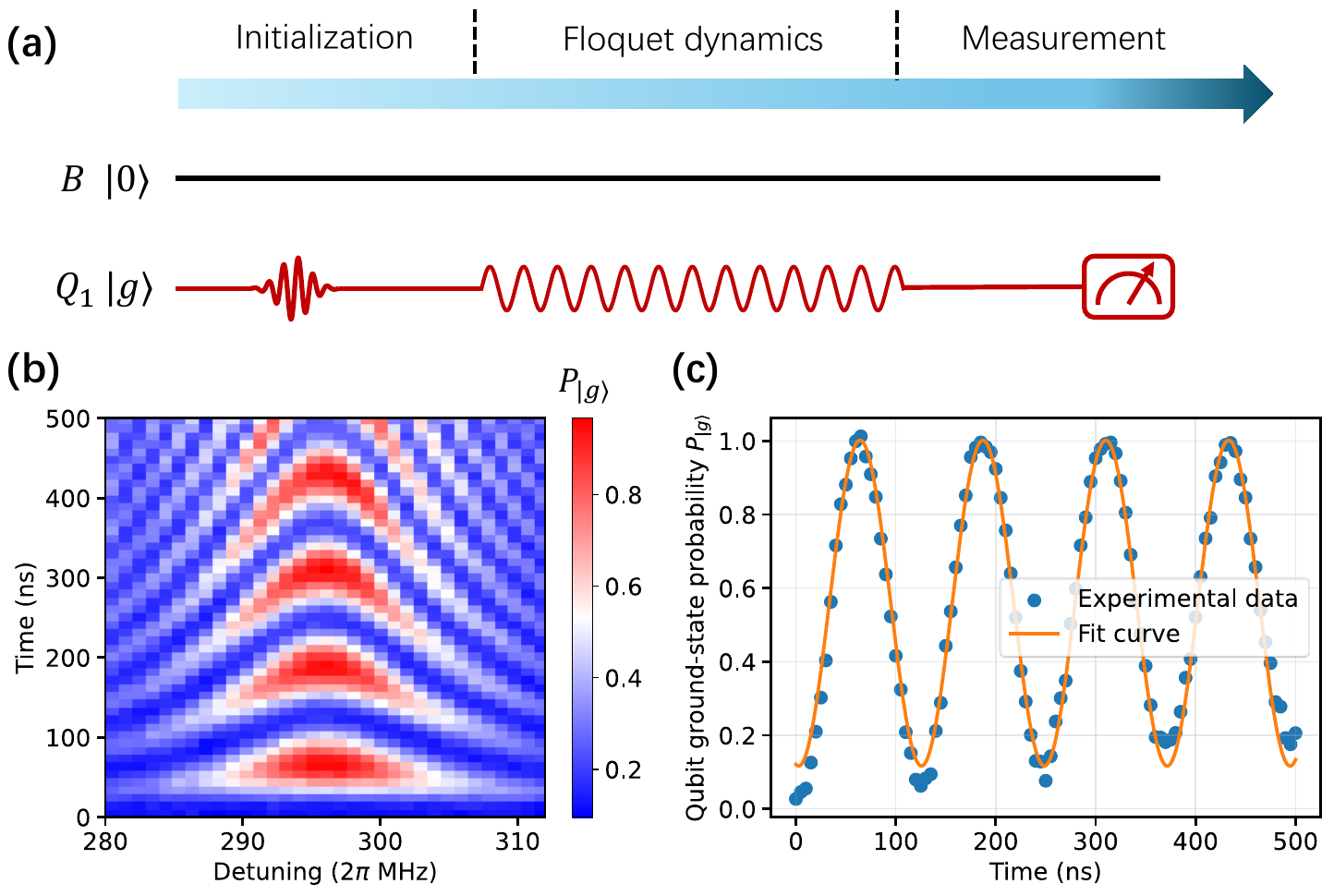}
    \caption{\textbf{Experimental demonstration of sideband coupling via Floquet modulation.}
    (a) Pulse sequence: a microwave \(\pi\)-pulse prepares the joint state \(|e0\rangle\) (qubit in \(|e\rangle\), cavity in vacuum \(|0\rangle_{\text{cav}}\)), followed by a Floquet \(Z\)-modulation and measurement.
    (b) Qubit ground-state probability \(P_g\) versus detuning \(\Delta_1\) and pulse duration under a Floquet pulse with \(\nu_1/(2\pi)=190\,\mathrm{MHz}\) and \(\varepsilon_1=2\pi \times 81.5\,\mathrm{MHz}\); the optimal operating point occurs near \(\Delta_1^{\mathrm{opt}}=2\pi \times 296\,\mathrm{MHz}\).
    (c) \(P_g\) versus pulse duration at \(\Delta_1^{\mathrm{opt}}\), showing coherent exchange \(|e0\rangle \leftrightarrow |g1\rangle\). A sinusoidal fit yields \(f_{\mathrm{swap}}=8.1\,\mathrm{MHz}\), implying an effective coupling \(\lambda_1/2=2\pi \times 4.1\,\mathrm{MHz}\).}
    \label{fig:FloquetExp}
\end{figure}

To verify the sideband coupling via Floquet modulation, we perform an experiment using a single qubit \(Q_1\) coupled to a cavity. The experimental setup and results are summarized in Fig.~\ref{fig:FloquetExp}. After initialization, a microwave \(\pi\)-pulse excites qubit \(Q_1\) from \(|g\rangle\) to \(|e\rangle\), preparing the joint state \(|e0\rangle = |e\rangle\!\otimes\!|0\rangle_{\text{cav}}\), where the trailing numeral denotes the cavity Fock state (\(0\) for vacuum, \(1\) for one photon). A Floquet \(Z\)-modulation with frequency \(\nu_1/(2\pi)=190\,\mathrm{MHz}\) and amplitude \(\varepsilon_1=2\pi \times 81.5\,\mathrm{MHz}\) is then applied, and we measure the qubit ground-state probability \(P_g\). Under this drive the system exhibits coherent exchange \(|e0\rangle \leftrightarrow |g1\rangle\), i.e., between the qubit excitation and a single cavity photon.

Figure~\ref{fig:FloquetExp}(b) shows \(P_g\) versus the detuning \(\Delta_1\) and pulse duration; the interference fringes reveal the sideband dynamics. From this map we identify the optimal detuning as \(\Delta_1^{\mathrm{opt}}=2\pi\times296\,\mathrm{MHz}\). The calibration of qubit frequency versus ZPA is performed under square-wave pulses; in the Floquet configuration this mapping is nonlinearly distorted, shifting the apparent optimum. In our device the qubit idle frequency is \(\omega_i/(2\pi)=5.3\,\mathrm{GHz}\) and the cavity frequency is \(\omega_s/(2\pi)=5.796\,\mathrm{GHz}\). For the first-order sideband, the ideal detuning predicted by the model is \(\Delta_{\mathrm{ideal}}=\omega_s-\omega_{i}-\nu_1=2\pi\times306\,\mathrm{MHz}\), which is slightly larger than the experimental \(\Delta_1^{\mathrm{opt}}\), consistent with the nonlinear ZPA mapping.

At \(\Delta_1=\Delta_1^{\mathrm{opt}}\), \(P_g\) versus Floquet pulse length [Fig.~\ref{fig:FloquetExp}(c)] oscillates with frequency \(f_{\mathrm{swap}}=8.1\,\mathrm{MHz}\), demonstrating coherent exchange \(|e0\rangle \leftrightarrow |g1\rangle\); this corresponds to an effective sideband coupling \(\lambda_1/2=2\pi\times4.1\,\mathrm{MHz}\). These observations quantitatively confirm the realization of sideband coupling via Floquet modulation in our system.

To benchmark the Floquet sideband across the full register, we extract for each qubit \(Q_j\) (\(j=1,\dots,8\)) the Floquet modulation frequency \(\nu_j/(2\pi)\), the effective coupling \(\lambda_j/2\), the modulation amplitude $\varepsilon_j$ from datasets analogous to Figs.~\ref{fig:FloquetExp}(b,c). The summary is organized in Table~\ref{tab:FloquetParams}.

\begin{table}[t]
\begin{center}
\setlength{\tabcolsep}{6pt}
\begin{tabular}{lcccccccc}
\hline\hline
Parameters & $Q_1$ & $Q_2$ & $Q_3$ & $Q_4$ & $Q_5$ & $Q_6$ & $Q_7$ & $Q_8$ \\
\hline
$\nu_j/(2\pi)$ (MHz) & 190 & 200 & 170 & 180 & 220 & 210 & 130 & 160 \\
$\lambda_j/2$ ($2\pi$ MHz) & 4.1 & 3.3 & 2.2 & 2.6 & 2.7 & 2.5 & 2.0 & 3.2 \\
$\varepsilon_j$ ($2\pi$ MHz) & 81.5 & 67.3 & 47.3 & 46.7 & 62.4 & 51.6 & 40.9 & 64.1 \\
    & & & & & & & & \\
\hline\hline
\end{tabular}
\end{center}
\caption{ \textbf{Floquet sideband parameters for qubits $\bm{Q}_1$--$\bm{Q}_8$}. Each entry lists the modulation frequency $\nu_j/(2\pi)$ used for the first-order sideband, the effective coupling $\lambda_j/2$ extracted from swap oscillations \(|e0\rangle \leftrightarrow |g1\rangle\) as in Fig.~\ref{fig:FloquetExp}(c), and the modulation amplitude $\varepsilon_j$.}
\label{tab:FloquetParams}
\end{table}

\section{Photon Number Distribution Measurement and Wigner Tomography}
\label{sec:pn-wigner}

\subsection{Photon-number readout via a resonant ancilla qubit}

\begin{figure}[H]
\centering
\includegraphics[width=0.95\linewidth]{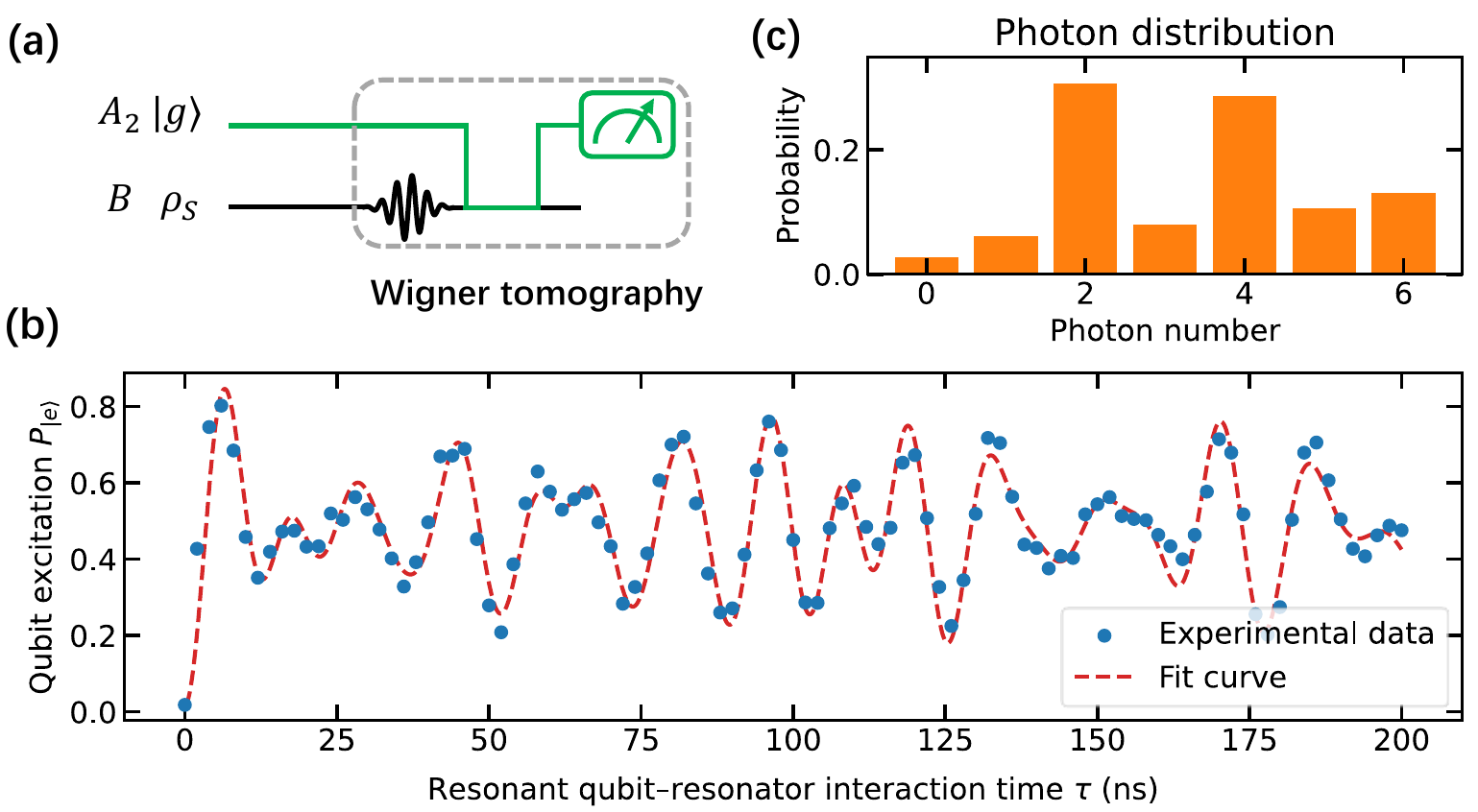}
\caption{\textbf{Photon-number readout and Wigner tomography.}
(a) Measurement sequence. The ancilla qubit $A_2$ is initialized in $\ket{g}$, while the bus resonator $B$ is in a state $\rho_s$ observed after dynamical evolution. After turning off all drives, $A_2$ is tuned on resonance with $B$ for a variable interaction time $\tau$, then parked back to its idle frequency and measured.
(b) Measured excited-state probability $P_e(\tau)$ of $A_2$ (blue dots) as a function of the resonant interaction time $\tau$. The red dashed curve is a least-squares fit to Eq.~\eqref{eq:Pe-sum}, where the oscillation frequencies $2\xi\sqrt{n}$ reflect contributions from different photon numbers.
(c) Photon-number distribution $\{P_n\}$ extracted from (b). The resonator state here is $\rho_s$ without any prior displacement $D(\beta)$, i.e., $\beta=0$. This example illustrates the case of a low-photon-number state.}
\label{fig:wigner-rabi}
\end{figure}

We use the ancilla qubit $A_2$ to read out the photon-number distribution of the bus resonator $B$. After a given dynamical evolution, the resonator reaches a state $\rho_s$. Figure~\ref{fig:wigner-rabi}(a) illustrates the measurement sequence: $A_2$ is initialized in $\ket{g}$, while the resonator is in $\rho_s$. After turning off all drives, the frequency of $A_2$ is tuned on resonance with $B$ for a variable interaction time $\tau$, after which the qubit is parked back to its idle point and measured.

The dynamics are governed by the resonant Jaynes--Cummings Hamiltonian
\begin{equation}
H=\omega_s a^\dagger a+\frac{\omega_s}{2}\sigma_z+\xi\,(a\,\sigma_+ + a^\dagger \sigma_-),
\end{equation}
where $a$ is the annihilation operator of $B$, $\sigma_{\pm}$ act on $\{\ket{g},\ket{e}\}$ of $A_2$, and $\xi$ is the coupling strength. In each Fock subspace $\{\ket{g,n},\ket{e,n-1}\}$ ($n\ge 1$), the system undergoes Rabi oscillations at frequency $2\xi\sqrt{n}$, while the vacuum $\ket{g,0}$ does not oscillate.

Let $P_n=\bra{n}\rho_s\ket{n}$ denote the photon-number distribution of the resonator. The excited-state probability of $A_2$ is then described by \cite{Hofheinz2009, Zheng2023PRL_PhaseTransition}
\begin{equation}
\label{eq:Pe-sum}
P_e(\tau)=\frac{1}{2}\Bigl\{\,1-\Bigl[P_{\ket{g}}(0)-P_{\ket{e}}(0)\Bigr]
\sum_{n=0}^{n_{\max}} P_n \cos\!\left(2\xi\sqrt{n}\,\tau\right)\Bigr\}.
\end{equation}
Here $P_{\ket{g}}(0)$ and $P_{\ket{e}}(0)$ denote the initial ground- and excited-state populations of $A_2$. In principle $P_{\ket{e}}(0)=0$, but in practice a small residual excitation is observed due to the finite detuning ($\sim700$ MHz) between the idle qubit frequency and the resonator when many photons are present. This offset causes the oscillations to center around the actual experimental mean excitation probability rather than exactly $1/2$.

The photon-number distribution $\{P_n\}$ is obtained by fitting the measured $P_e(\tau)$ to Eq.~\eqref{eq:Pe-sum}. In practice, we perform a least-squares fit, where the free parameters are the probabilities $\{P_n\}$ constrained to satisfy $P_n\ge0$ and $\sum_n P_n=1$. An example of the measured Rabi signal and the extracted distribution is shown in Figs.~\ref{fig:wigner-rabi}(b,c).

 \paragraph{Choice of the resonant readout.}
For a representative dispersive shift $\chi/(2\pi)\simeq1$~MHz, a dispersive parity mapping requires about $0.5~\mu\mathrm{s}$ for phase accumulation, before the two Ramsey $\pi/2$ pulses and qubit readout are included. This is comparable to the cat-state decoherence time $2T_{1,s}/|\alpha|^2\simeq0.55~\mu\mathrm{s}$ for $T_{1,s}=3.0~\mu\mathrm{s}$ and $|\alpha|=3.3$. The resonant Jaynes--Cummings record used here lasts about $200$~ns and provides the displaced photon-number distribution, from which the parity is obtained as $\sum_n(-1)^nP_n$.

\begin{figure}[p]
\centering
\includegraphics[width=0.98\linewidth]{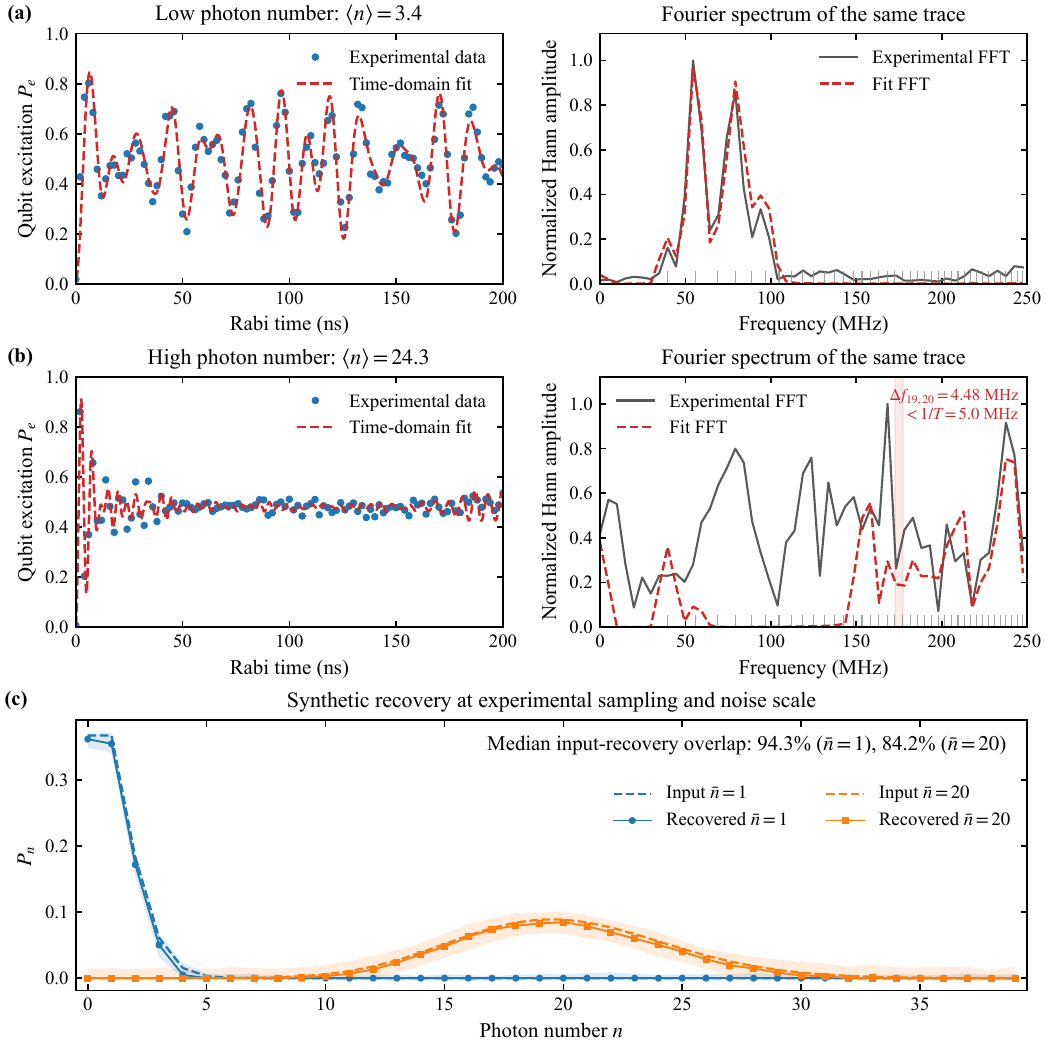}
\caption{\textbf{Time-domain fits, Fourier spectra, and synthetic recovery.} (a,b) Representative low- and high-photon-number records, respectively. The left panels show the measured qubit-excitation probability and constrained time-domain fit; the right panels show the Hann-window Fourier spectra of the measured and fitted traces, evaluated on the same 101 samples over 200~ns. The short gray ticks mark the expected Jaynes--Cummings frequencies $f_n=2[\xi/(2\pi)]\sqrt{n}$. The mean absolute residuals in $P_e$ are $0.040$ in (a) and $0.022$ in (b); the smaller value in (b) mainly reflects the weak fluctuations around $P_e=0.5$ after the initial transient and is not a measure of improved high-$n$ resolution. (c) Recovery of known Poisson photon-number distributions from synthetic traces with the experimental sampling and noise scale. Dashed curves are the inputs, solid curves and markers are the median recovered distributions, and shaded bands contain the 10th--90th percentiles over 200 noise realizations. The quoted overlap is $\sum_n\min(P_n^{\rm in},P_n^{\rm rec})$.}
\label{fig:rabi-low-high}
\end{figure}

\paragraph{Time-domain photon-number extraction and Fourier diagnostic.}
To illustrate how the information appears in the two domains, Fig.~\ref{fig:rabi-low-high} pairs the measured Rabi response and constrained time-domain fit with the Fourier spectrum of the same trace. The low-photon-number example in panel (a) is the trace already shown in Fig.~\ref{fig:wigner-rabi}(b), whereas panel (b) is taken from a strongly displaced point of the tomography grid, where $\langle n\rangle=24.3$. Both time-domain traces are reproduced by the fit. The smaller residual in panel (b) mainly reflects the fact that, after the initial transient, the signal remains close to $P_e=0.5$ with relatively small excursions; it does not mean that every individual high-$n$ weight is determined more accurately. This limitation is more transparent in the Fourier domain. The $200$-ns acquisition window sets a nominal frequency scale $1/T=5.0$~MHz, while the independently calibrated coupling $\xi/(2\pi)=19.8$~MHz gives a separation of only $4.48$~MHz between the $n=19$ and $20$ Jaynes--Cummings components. Thus adjacent high-$n$ components cannot be resolved as distinct Fourier peaks, even though their frequencies are known and can be used jointly in the time-domain model. We therefore use the Fourier transform as a diagnostic of the measured frequency content rather than extracting $P_n$ directly from individual peak heights. Panel (c) tests the inversion with known inputs: synthetic Rabi traces are generated on the same time grid and are then analyzed by the same constrained fit. Independent Gaussian noise is added to the synthetic traces, with its standard deviation set by the RMS residual of the experimental fits. The median overlap $\sum_n\min(P_n^{\rm in},P_n^{\rm rec})$ between the input and recovered distributions is $94.3\%$ for $\bar n=1$ and $84.2\%$ for $\bar n=20$. The recovery therefore retains the location and overall envelope of the distribution at high photon number, while becoming less precise for the individual weights.

  \subsection{Wigner tomography}

To reconstruct the resonator state in phase space, we apply a displacement operator
\begin{equation}
D( \beta )=\exp\!\big( \beta a^\dagger- \beta ^*a\big)
 \end{equation}
realized by a calibrated resonator drive whose amplitude and phase define $\beta$. The displaced state is
\begin{equation}
\tilde{\rho}_s( \beta )=D^\dagger( \beta )\,\rho_s\,D( \beta )
 \end{equation}

The photon-number probabilities of the displaced state are
\begin{equation}
\tilde{P}_n( \beta )=\bra{n}\tilde{\rho}_s( \beta )\ket{n}
 \end{equation}
The Wigner function is then obtained from the parity relation
\begin{equation}
W( \beta )=\frac{2}{\pi}\sum_{n=0}^{\infty}(-1)^n\tilde{P}_n( \beta )
 \end{equation}

For convenience, we parameterize $\beta$ in terms of the canonical quadratures
\begin{equation}
x= \frac{\beta+\beta^*}{\sqrt{2}},\qquad
p= \frac{\beta-\beta^*}{i\sqrt{2}}
 \end{equation}
so that $\beta=(x+ip)/\sqrt{2}$. By scanning $\mathrm{Re}\,\beta$ and $\mathrm{Im}\,\beta$ over a two-dimensional grid, we reconstruct $W(\mathrm{Re}\,\beta,\mathrm{Im}\,\beta)$ of the state $\rho_s$, as shown in the main text.

\begin{figure*}[t]
  \centering
   \includegraphics[width=0.82\linewidth]{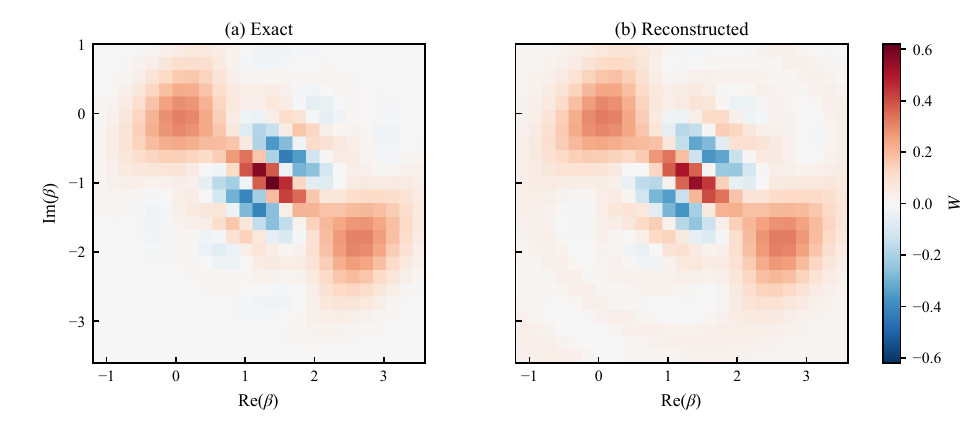}
 \caption{\textbf{Unitary three-level check of the resonant Wigner readout.} (a) Exact Wigner function of the synthetic truncated cat. (b) Wigner function reconstructed from the three-level simulated Rabi signals using the two-level procedure applied to the data. Both panels use the same color scale.}
  \label{fig:qutrit-wigner-comparison}
\end{figure*}

    To quantitatively examine the validity of the readout procedure, we generate Rabi signals with a three-level Jaynes--Cummings model that takes the second excited state of the ancilla qubit into consideration. The binary excited outcome is modeled as $P_{\rm exc}=P_e+P_f$. We then extract the photon-number distributions using the same two-level fitting procedure as for the data and obtain the Wigner function from their parity. Figure~\ref{fig:qutrit-wigner-comparison} compares the reconstructed Wigner function with the exact result and shows agreement in the main phase-space structure. Relaxation and dephasing are omitted here to isolate the coherent multilevel dynamics.

\begin{figure*}[t]
\centering
\includegraphics[width=\linewidth]{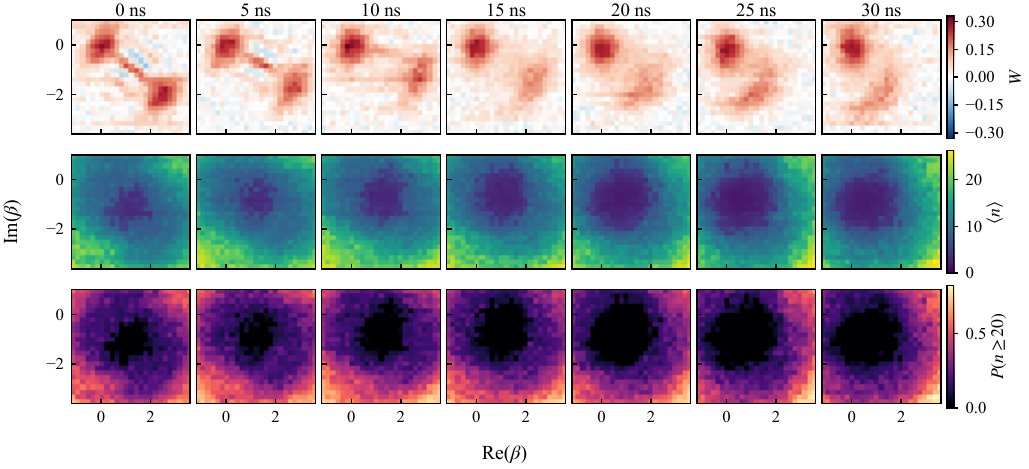}
\caption{\textbf{Photon-number content of the Wigner tomography grid.} Top: measured Wigner functions for the seven interaction times in main-text Fig.~3(b). Middle: displaced mean photon number $\langle n\rangle$. Bottom: fitted probability $P(n\geq20)$. The largest photon numbers occur predominantly at the outer phase-space corners, whereas the cat-state lobes and interference corridor occupy lower-photon-number regions.}
\label{fig:wigner-photon-number}
\end{figure*}

 A characteristic photon-number scale for these multilevel effects can be estimated from the resonant Jaynes--Cummings oscillation. For an initial Fock component $|g,n\rangle$, the two-level model gives

\begin{equation}
P_e^{(n)}(t)=\sin^2\!\left(\xi\sqrt{n}\,t\right),\qquad
\Omega_n=2\xi\sqrt{n},
\label{eq:multilevel-photon-scale-rabi}
\end{equation}
where $\Omega_n$ is the photon-number-dependent Rabi angular frequency. The $|e\rangle\leftrightarrow|f\rangle$ transition is separated from the resonant $|g\rangle\leftrightarrow|e\rangle$ transition by the ancilla anharmonicity $K=\omega_{ef}-\omega_{ge}$. A simple scale for the onset of multilevel effects is therefore obtained by comparing $\Omega_n$ with $|K|$:

\begin{equation}
2\xi\sqrt{n_c}\sim|K|,\qquad
n_c\sim\left(\frac{|K|}{2\xi}\right)^2
=\left(\frac{250}{2\times19.8}\right)^2
=39.9\simeq40.
\label{eq:multilevel-photon-scale}
\end{equation}
Here $|K|/(2\pi)\simeq250$~MHz and $\xi/(2\pi)=19.8$~MHz. This comparison gives an order-of-magnitude scale rather than a sharp threshold; the full three-level simulation in Fig.~\ref{fig:qutrit-wigner-comparison} provides the quantitative check. Figure~\ref{fig:wigner-photon-number} further shows that photon numbers above $20$ occur mainly at the outer corners of the tomography grid, while the cat-state lobes and interference region involve considerably fewer photons.

\pagebreak

 \section{Quantification of the distinguishability}

\subsection{Definition}
The distinguishability $\mathcal{D}$ quantifies the degree to which the reservoir states corresponding to different field branches can be distinguished. It is defined as the trace distance
\begin{equation}
    \mathcal{D} = \frac{1}{2}\,\mathrm{Tr}\left|\rho^{|0\rangle}-\rho^{|\alpha\rangle}\right|,
\end{equation}
where $\rho^{|0\rangle}$ and $\rho^{|\alpha\rangle}$ are the reduced reservoir states conditioned on the vacuum and coherent components of the field, respectively. The notation $|\cdot|$ indicates the positive semidefinite operator obtained by taking absolute values of the eigenvalues, i.e.,
\begin{equation}
    \mathcal{D}=\frac{1}{2}\sum_i |\zeta_i|
\end{equation}
with $\{\zeta_i\}$ the eigenvalues of $\Delta=\rho^{|0\rangle}-\rho^{|\alpha\rangle}$.

\subsection{Derivation of the distinguishability}

Consider an imperfect equal-weight cat state described by
\begin{equation}
\rho_s(0)=\frac{1}{2}\left[
|0\rangle\langle0|+|\alpha\rangle\langle\alpha|
+\eta|\alpha\rangle\langle0|+\eta^*|0\rangle\langle\alpha|
\right].
\label{eq:imperfect-cat-initial}
\end{equation}
Let $|G\rangle=\bigotimes_{k=1}^N|g\rangle_k$. Under the joint unitary evolution, $|0,G\rangle$ remains unchanged and $|\alpha,G\rangle$ evolves to $|\psi_\alpha(t)\rangle$. The total density operator is therefore
\begin{equation}
\rho(t)=\frac{1}{2}\left[
|0,G\rangle\langle0,G|
+|\psi_\alpha\rangle\langle\psi_\alpha|
+\eta|\psi_\alpha\rangle\langle0,G|
+\eta^*|0,G\rangle\langle\psi_\alpha|
\right].
\label{eq:imperfect-cat-evolved}
\end{equation}
The state $|\psi_\alpha\rangle$ can be expanded as
\begin{equation}
|\psi_\alpha\rangle=\sum_{n=0}^{\infty}d_n|n\rangle|\phi_n^q\rangle.
\end{equation}
Tracing out the cavity then gives
\begin{equation}
\rho_q(t)\simeq
\frac{1}{2}\left(\rho^{|0\rangle}+\rho^{|\alpha\rangle}\right),
\quad
\rho^{|0\rangle}=|G\rangle\langle G|,
\quad
\rho^{|\alpha\rangle}=\sum_{n\ge1}|d_n|^2|\phi_n^q\rangle\langle\phi_n^q|.
\label{eq:cat-reservoir-mixture}
\end{equation}
Here the vacuum amplitude in $|\psi_\alpha\rangle$ has been neglected; at the experimental cat size it is small. Within the timescale $\pi/(|\alpha|\lambda_k)$, the correlation between the qubits can be neglected so that $\rho^{|\alpha\rangle}$ can be written as $\rho^{|\alpha\rangle}=\bigotimes_{k=1}^{N}\rho_k^{|\alpha\rangle}$. Consequently, the reduced density operator of the $k$th qubit, obtained by tracing out the states of all other qubits in $\rho_q(t)$, is given by
\begin{equation}
\rho_k=\frac{1}{2}\left(|g\rangle_k\langle g|+\rho_k^{|\alpha\rangle}\right),
\end{equation}
which leads to $\rho_k^{|\alpha\rangle}=2\rho_k-|g\rangle_k\langle g|$.

\subsection{Ensuring Positive Semidefinite Reconstruction}

The approximate single–qubit branch state obtained from tomography,
\begin{equation}
\label{eq:rho-alpha-raw}
\rho_k^{|\alpha\rangle} \simeq 2\,\rho_k - \ket{g}_k\bra{g},
\end{equation}
provides a convenient linear estimate. However, the linear relation in Eq.~\eqref{eq:rho-alpha-raw} does not, in general, preserve positive semidefiniteness, even in the absence of experimental noise. To restore physicality, $\rho_k^{|\alpha\rangle}$ is projected onto the set of unit-trace positive semidefinite matrices via eigenvalue decomposition.

We first compute its spectral decomposition,
\begin{equation}
\label{eq:eigh}
\rho_k^{|\alpha\rangle}
\;=\;
V\,\mathrm{diag}(\zeta_1,\ldots,\zeta_d)\,V^{\dagger},
\end{equation}
and then remove any negative eigenvalues,
\begin{equation}
\label{eq:clip}
\zeta_i' \;=\; \max\!\bigl(\zeta_i,\,0\bigr)
\qquad (i=1,\ldots,d).
\end{equation}
The matrix is subsequently normalized to restore unit trace,
\begin{equation}
\label{eq:renorm}
\tilde{\rho}_k^{|\alpha\rangle}
\;=\;
V\,\mathrm{diag}\!\Bigl(\frac{\zeta_1'}{\sum_{j}\zeta_j'},\ldots,\frac{\zeta_d'}{\sum_{j}\zeta_j'}\Bigr)\,V^{\dagger}.
\end{equation}
The resulting matrix satisfies
\begin{equation}
\label{eq:constraints}
\tilde{\rho}_k^{|\alpha\rangle}\ \text{is positive semidefinite},
\qquad
\mathrm{Tr}\!\bigl[\tilde{\rho}_k^{|\alpha\rangle}\bigr]=1,
\end{equation}
and coincides with Eq.~\eqref{eq:rho-alpha-raw} whenever the latter is already positive semidefinite and normalized.

Applying this procedure qubit by qubit ensures that all reconstructed branch states are physical, and that the factorized reservoir state
\begin{equation}
\label{eq:manyqubit}
\rho^{|\alpha\rangle}\;\simeq\;\bigotimes_{k=1}^{N}\tilde{\rho}_k^{|\alpha\rangle}
\end{equation}
remains a valid positive semidefinite density operator for further analysis.

   \subsection{Direct experimental validation with separately prepared branches}
\label{sec:D-truncated-benchmark}

\begin{figure*}[t]
\centering
\includegraphics[width=0.96\linewidth]{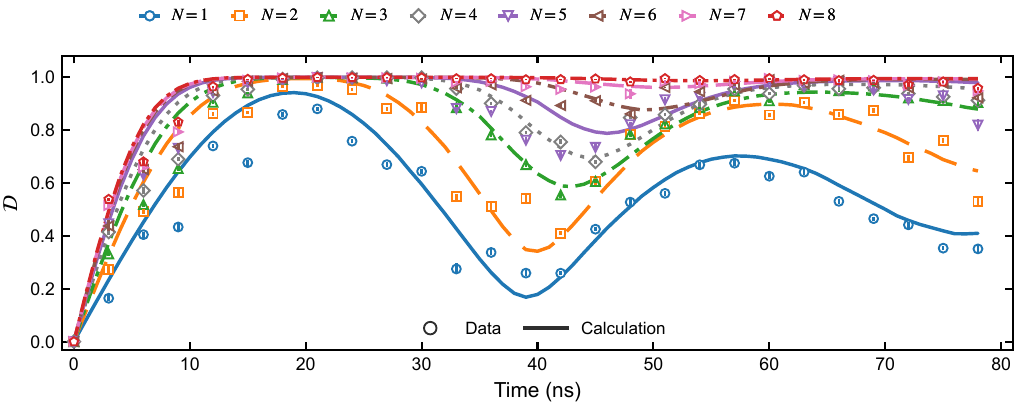}
\caption{\textbf{Direct-branch check of the distinguishability reconstruction.} Markers and error bars show the product-of-marginals trace distance obtained by separately preparing $|0\rangle$ and $|\alpha\rangle$; curves show the corresponding calculations.}
\label{fig:D-truncated-validation}
\end{figure*}

 We here obtain the distinguishability by separately preparing $|0\rangle$ and $|\alpha\rangle$ for each reservoir size $N=1,\ldots,8$, measuring the single-qubit density matrices, and evaluating the same product-of-marginals trace distance:
\begin{equation}
\mathcal D_{\mathrm{direct,prod}}(t)
=\frac{1}{2}\left\|
\bigotimes_{k=1}^{N}\rho_k^{|0\rangle}(t)
-
\bigotimes_{k=1}^{N}\rho_k^{|\alpha\rangle}(t)
\right\|_1.
\label{eq:D-direct-product}
\end{equation}
Figure~\ref{fig:D-truncated-validation} shows the resulting direct-branch measurements and the corresponding calculations. The markers show the mean and standard deviation of ten repetitions, and the common initial reservoir state gives $\mathcal D(0)=0$. The curves are obtained by propagating the two branches and applying the same product-of-marginals analysis.

In the calculations, the amplitude of the coherent state is set to be $\alpha=3.3$, and the pulse envelope and decoherence effects are taken into consideration. Small parameter adjustments account for calibration drift. The distinguishabilities directly obtained from these two separate experiments are in good agreement with the results displayed in Fig.~3(a) of the main text, confirming the progressive suppression of the distinguishability revival with the increase of $N$.

 \clearpage
  \subsection{Dependence on the initial cat state}
\label{sec:initial-state-comparison}

\begin{figure}[H]
\centering
\includegraphics[width=0.97\linewidth]{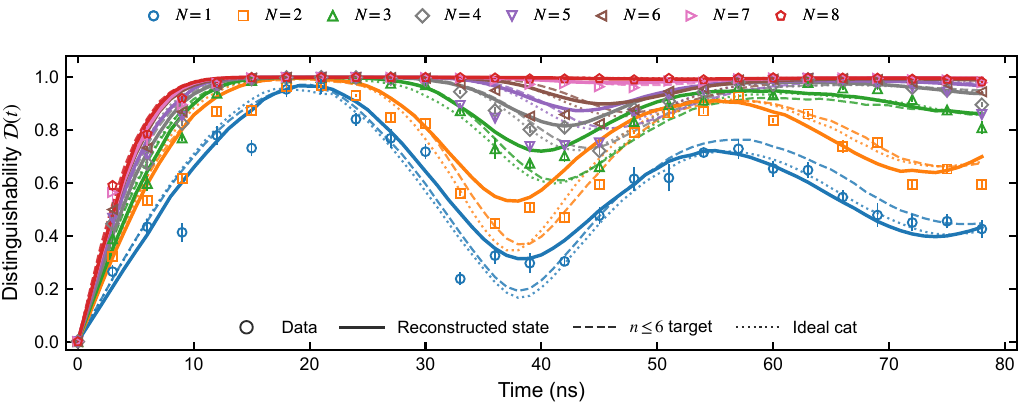}
\caption{\textbf{Dependence of the calculated distinguishability on the initial cat state.} Markers and error bars show the experimental data. For $N=1$--8, solid lines use the independently reconstructed experimental amplitude-cat density matrix, dashed lines use the ideal $n\leq6$ truncated target state, and dotted lines use an ideal amplitude cat with a slightly adjusted coherent amplitude. The same dynamical parameters and product-of-marginals analysis are otherwise used in all three calculations.}
\label{fig:initial-state-comparison}
\end{figure}

 We compare the independently reconstructed experimental amplitude-cat density matrix used in main-text Fig.~3(a) with two idealized descriptions: the ideal $n\leq6$ truncated target state and an ideal amplitude cat. For the ideal-cat calculation, the coherent amplitude is slightly adjusted from its nominal value to account for calibration drift, while all other dynamical parameters are kept unchanged. Figure~\ref{fig:initial-state-comparison} also includes the experimental data for reference. The three calculations show modest differences around the revivals, while all give the same progressive suppression as $N$ increases.

\section{Local single-qubit entropies of the reservoir}
\label{sec:reservoir-entropy-correlations}

\begin{figure}[H]
\centering
\includegraphics[width=\linewidth]{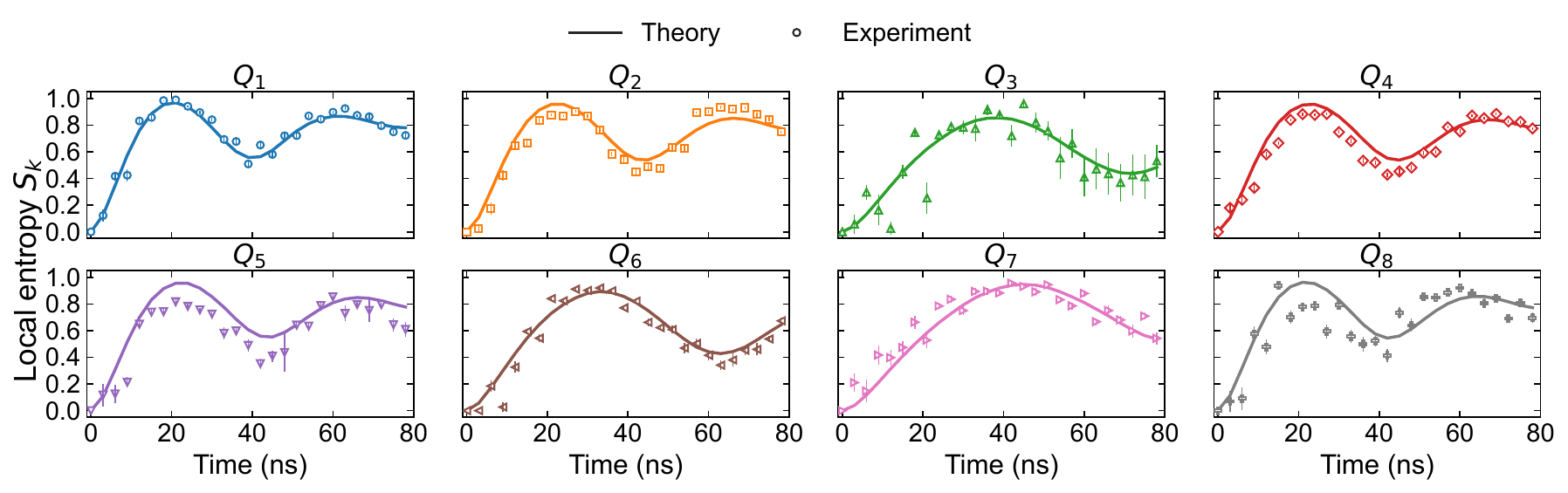}
\caption{\textbf{Local entropies of the $N=8$ reservoir.} Markers show the mean values obtained from single-qubit tomography over ten repetitions; error bars denote one standard deviation. Solid curves are Monte Carlo wave-function simulations including resonator and qubit relaxation and dephasing.}
\label{fig:reservoir-entropy-correlations}
\end{figure}

For each reservoir qubit, we measure the three Bloch-vector components and reconstruct its reduced state as
\begin{equation}
\rho_k=\frac{1}{2}\left(I+\langle\sigma_x\rangle\sigma_x+\langle\sigma_y\rangle\sigma_y+\langle\sigma_z\rangle\sigma_z\right).
\end{equation}
The corresponding entropy is
\begin{equation}
S_k=-\mathrm{Tr}(\rho_k\log_2\rho_k).
\end{equation}
Figure~\ref{fig:reservoir-entropy-correlations} shows the evolutions of the 8 local single-qubit entropies measured for $N=8$. The rise and partial decrease of each local entropy are consistent with the transfer and partial return of which-path information during the interaction. Because the qubits have different effective couplings, their entropy oscillations are not synchronized. The solid curves are Monte Carlo wave-function simulations of the Tavis--Cummings dynamics, including the measured relaxation and dephasing of the resonator and all eight qubits. The effective coupling strengths are fine-tuned to account for calibration drift.

The local qubit entropies characterize the entanglement between individual qubits and the field mode. However, when $N>1$, neither these local entropies nor their sum can be used as a measure to quantify the amount of which-path information stored in the reservoir, which is determined by the distance between the collective states of the qubits correlated to the two field components.

 \clearpage

\section{Interaction-induced deformation of the Wigner function}
\label{sec:JC-Wigner-deformation}

To examine the deformation observed in main-text Fig.~3(b), we evolve the cavity state reconstructed from the measured $0$-ns Wigner function with the effective eight-qubit Jaynes--Cummings model, written in frequency units as

\begin{equation}
\mathcal{H}_{\mathrm{JC}}
=\sum_{j=1}^{8}\delta_j |e_j\rangle\langle e_j|
+\sum_{j=1}^{8}\frac{\lambda_j}{2}
\left(a\sigma_j^+ +a^\dagger\sigma_j^-\right).
\label{eq:multiqubit-JC-Wigner}
\end{equation}
The effective couplings $\lambda_j/2$ and detunings $\delta_j$ are the calibrated parameters used for the Fig.~3 calculations. All reservoir qubits are initialized in $|g\rangle$. After the interaction, the qubits are traced out and the cavity Wigner function is calculated from the resulting reduced density matrix. No separate cavity Kerr term is included in Eq.~\eqref{eq:multiqubit-JC-Wigner}.

\begin{figure}[H]
\centering
\includegraphics[width=\linewidth]{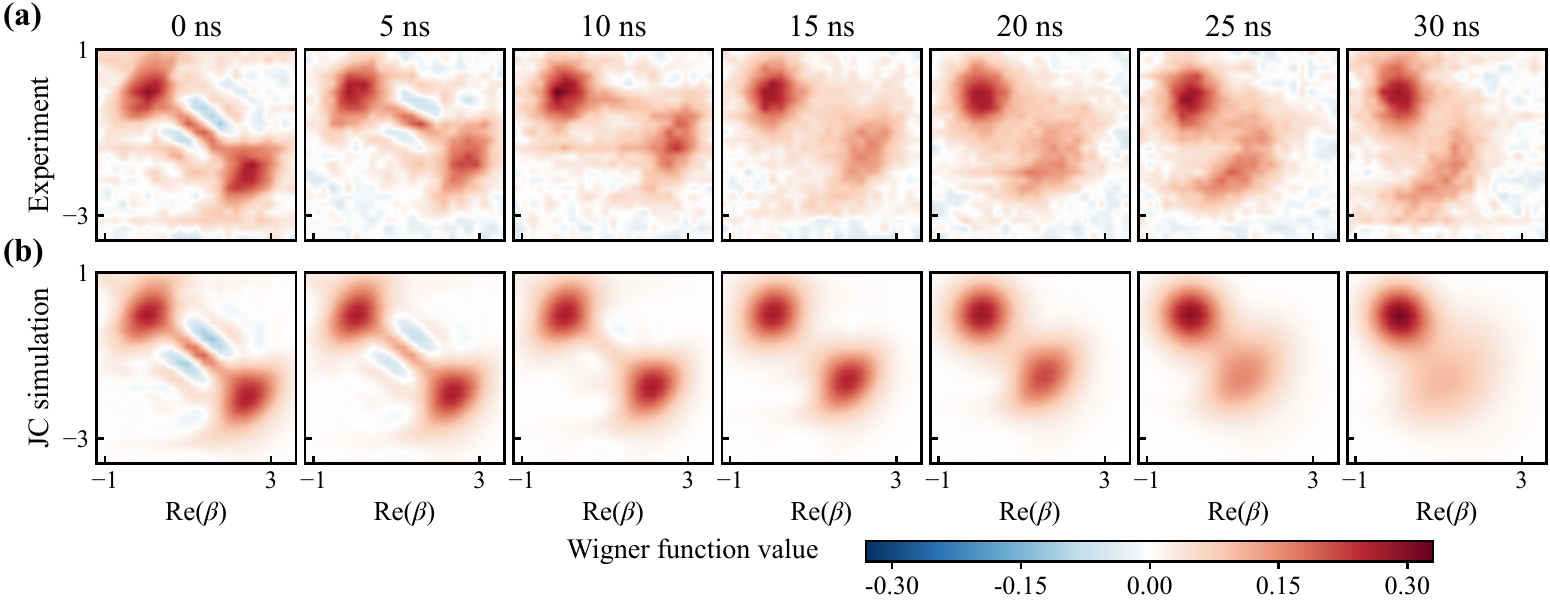}
\caption{\textbf{Interaction-induced deformation of the Wigner function.} (a) Experimental Wigner functions from main-text Fig.~3(b). (b) Wigner functions obtained by propagating the reconstructed $0$-ns cavity state under the effective eight-qubit Jaynes--Cummings Hamiltonian. Both rows are shown in the same phase-space coordinates and with the same symmetric color scale; white corresponds to $W=0$.}
\label{fig:JC-Wigner-deformation}
\end{figure}

As shown in Fig.~\ref{fig:JC-Wigner-deformation}, the multiqubit Jaynes--Cummings dynamics produces the phase spreading and deformation of the coherent-state component. Although no separate cavity Kerr term is included, the photon-number-dependent dressed-state dynamics gives rise to the self-Kerr effect inherited from the interaction with the qubits \mbox{
\cite{Auffeves2003PRL}}\hskip0pt
.

\section{Imperfect revival}

\begin{figure}[t]
    \centering
    \includegraphics[width=\linewidth]{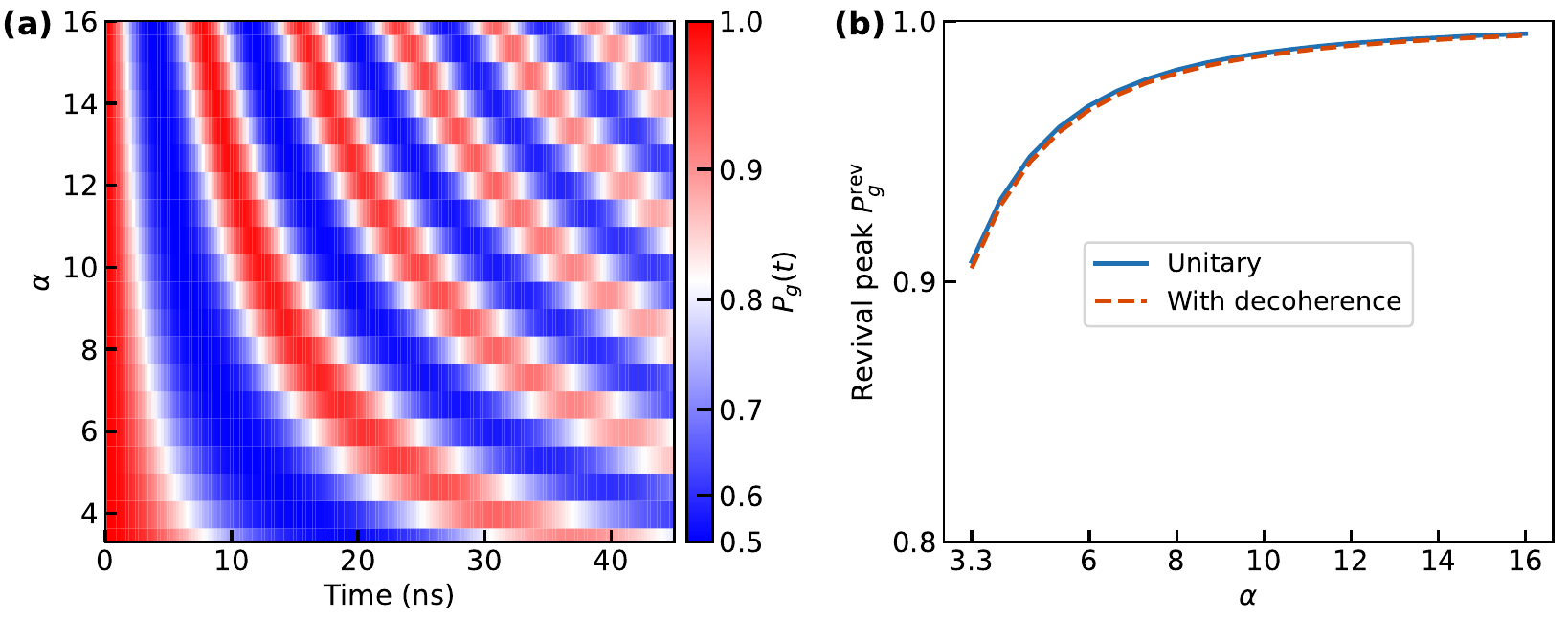}
    \caption{
    \textbf{Imperfect revival and its weak sensitivity to relaxation and dephasing.}
    (a) Density plot of the qubit ground-state population $P_g(t)$ versus time and coherent-state amplitude $\alpha$.
    The revival becomes more complete as $\alpha$ increases, including around the experimental value $\alpha=3.3$.
    (b) Extracted revival peak $P_g^{\mathrm{rev}}$ versus $\alpha$, comparing unitary evolution and the case including decoherence (energy relaxation and dephasing).
    The open-system simulation uses the measured $T_1$ and $T_2$ of the reservoir qubit ($Q_1$), and cavity decoherence parameters from Sec.~\ref{QR_Parameters}.
    }
    \label{fig:S7-imperfect-revival}
\end{figure}

Figure~\ref{fig:S7-imperfect-revival}(a) shows the qubit ground-state population $P_g(t)$ as a function of time $t$ and the coherent-state amplitude $\alpha$, including the experimental value $\alpha=3.3$.
The revival of $P_g(t)$ is not perfectly complete, and this imperfection becomes less pronounced as $\alpha$ increases.
This behavior can be attributed to the photon-number-dependent Rabi frequencies associated with different Fock components in $|\alpha\rangle$ (and the corresponding cat state), which results in a mild dephasing among these components.

To verify that the imperfect revival is not primarily caused by relaxation and dephasing, we additionally simulate the open-system dynamics by including decoherence (energy relaxation and dephasing) with finite $T_1$ and $T_2$.
In these simulations, the reservoir qubit is assigned the measured $T_1$ and $T_2$ values of the reservoir qubit ($Q_1$), while the cavity decoherence parameters are taken from Sec.~\ref{QR_Parameters}.
Figure~\ref{fig:S7-imperfect-revival}(b) summarizes the revival quality by plotting the peak value of the first revival, $P_g^{\mathrm{rev}}$, as a function of $\alpha$, comparing unitary evolution and evolution with decoherence.
The revival peak increases with $\alpha$ and approaches unity, while the difference between the unitary and open-system cases remains small in this parameter range.

\begingroup
\section{Observation of non-Markovianity}

Going one step further, we investigate how the degree of non-Markovianity of the decoherence dynamics depends upon the number of the reservoir qubits. The measure of Ref. \cite{Breuer2009PRL} is defined as the time-integration of the changing rate of the distinguishability between a pair of states over all time intervals during which the changing rate is positive, maximized over all pairs of initial states of the system. For an infinite-dimensional system, it is a challenging task to single out such a pair of initial states. Furthermore, the changing rate may keep positive even after a very long evolution time, and the result may become divergent. We here introduce a new measure, defined as the decrease of the which-path information leaking to the reservoir,
\begin{equation}
{\cal N}_{m}=\max_{[t_{i},\tau_{i}]}\int_{t_{i}}^{\tau_{i}}dt\left|\sigma\right|,
\end{equation}
maximized over all time intervals, $[t_{i},\tau_{i}]$, during which $\sigma=d{\cal D}/dt<0$. Here the distinguishability, given by Eq. (7) in the main text, is defined in terms of the reservoir states and is therefore different from that used in Ref. \cite{Breuer2009PRL}. This measure quantifies the maximal amount of which-path information that can be spontaneously erased as a consequence of energy backflow. When the cat state is used to encode quantum information \cite{Xu2020PRL}, ${\cal D}$ measures the loss of quantum information, while ${\cal N}_{m}$ quantifies the part that can return through the subsequent dynamics.

Within the measured time window, the dominant decreasing interval extends from the first maximum of ${\cal D}(t)$ to the following minimum. We use this same prescription to extract ${\cal N}_{m}$ from the cat-state data in main-text Fig.~3(a), the separately prepared branch data in Fig.~\ref{fig:D-truncated-validation}, and their corresponding calculations. Figure~\ref{fig:non-M} shows that the two experimental determinations and the calculations give the same progressive suppression of ${\cal N}_{m}$ as the reservoir size increases.

\begin{figure}[H]
    \centering
    \includegraphics[width=0.82\linewidth]{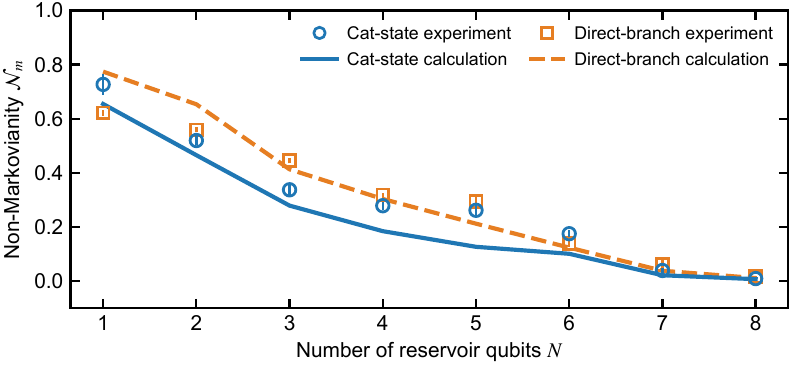}
    \caption{\textbf{Non-Markovianity obtained from cat-state inference and separately prepared branches.}
    Open markers show the experimental values of ${\cal N}_{m}$ extracted from the cat-state data in main-text Fig.~3(a) (circles) and the direct-branch data in Fig.~\ref{fig:D-truncated-validation} (squares); error bars are propagated from the corresponding distinguishability values. Curves show ${\cal N}_{m}$ extracted in the same way from the corresponding calculations.
    }
    \label{fig:non-M}
\end{figure}
\endgroup

\section{Numerical check of the field-state approximation}

\begin{figure}[H]
    \centering
    \includegraphics[width=0.9\linewidth]{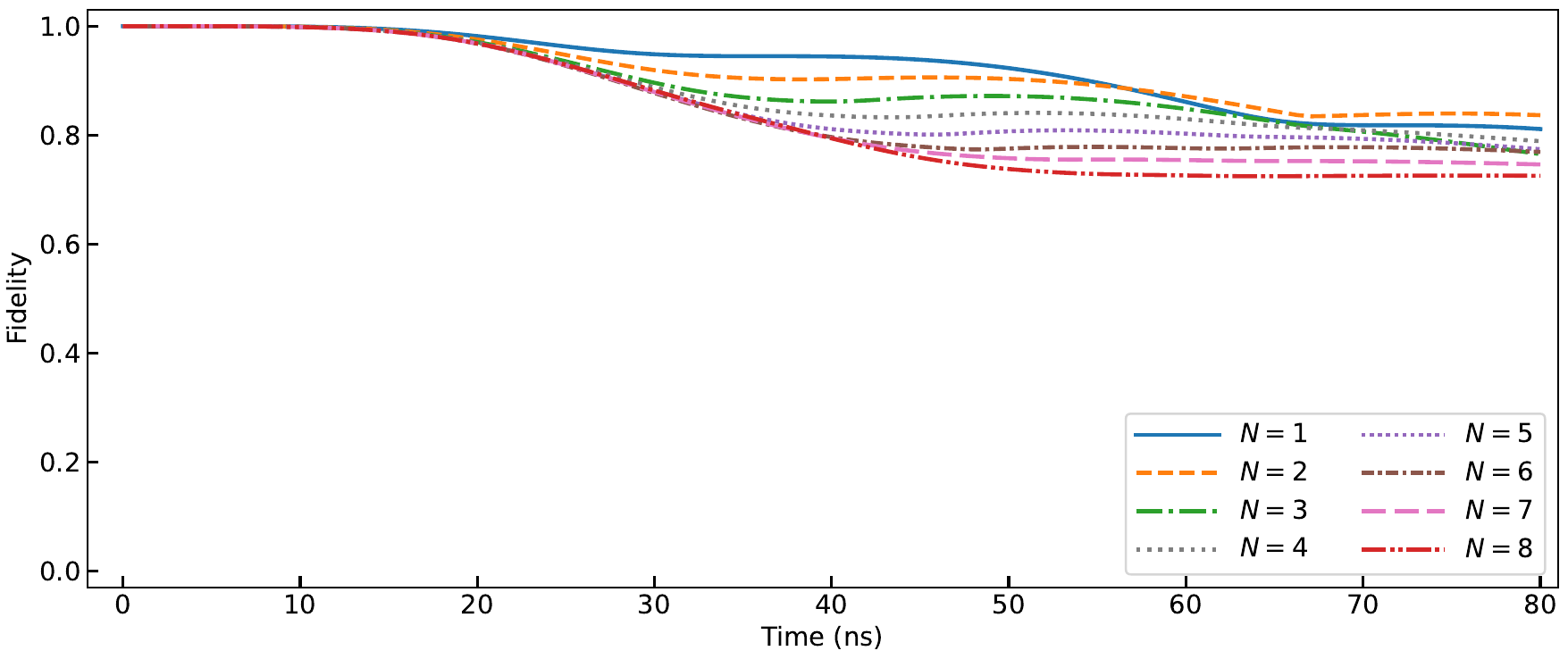}
    \caption{
    \textbf{Numerical check of the field-state approximation at the experimental cat size.}
    Time evolution of the fidelity for $\alpha=3.3$ and $N=1$ to $8$ reservoir qubits. Here $\rho_s(t)$ is obtained from main-text Eq.~(6), while $\rho'_s(t)$ is the reduced field state computed from the full evolution under main-text Eq.~(1) followed by tracing out the reservoir. The fidelity remains close to unity at early times and decays faster as $N$ increases.
    }
    \label{fig:S8-fidelity}
\end{figure}

To provide numerical evidence for the validity of the approximations in Eq.~(4) and Eq.~(6) of the main text, we simulate the reduced field dynamics for the experimental cat-state size $\alpha=3.3$.
We compare the field reduced density operator $\rho_s(t)$ obtained from the approximation with the reduced density operator $\rho'_s(t)$ obtained from the full evolution under the Hamiltonian of main-text Eq.~(1) followed by tracing out the reservoir qubits.
When constructing $\rho_s(t)$ from main-text Eq.~(6), we allow a small adjustment of the coherent-state amplitude $\alpha$ to incorporate the slight deformation of the coherent-state component induced by the qubit backaction.
The agreement is quantified using the Uhlmann fidelity,
\begin{equation}
\mathcal{F}(t)=\mathrm{Tr}\sqrt{\sqrt{\rho_s(t)}\,\rho'_s(t)\,\sqrt{\rho_s(t)}}.
\end{equation}
Figure~\ref{fig:S8-fidelity} shows $\mathcal{F}(t)$ for $N=1$ to $8$ reservoir qubits.
For the parameters used here, $\mathcal{F}(t)$ decreases faster as $N$ increases.
The deviation from unity mainly comes from a slight deformation of the coherent-state component $|\alpha\rangle$ induced by the backaction of the qubits.
Notably, during the which-path acquisition stage, this deformation remains weak, and the fidelity for each $N$ stays high before $\mathcal{D}$ reaches its maximum.

\clearpage